\documentclass[lettersize,journal]{IEEEtran}
\usepackage{cite}
\usepackage{amsmath,amssymb,amsfonts}
\usepackage{algorithmic}
\usepackage{graphicx}
\usepackage{textcomp}
\usepackage{caption}
\usepackage{subcaption}
\usepackage{float}
\usepackage[utf8]{inputenc}
\usepackage{bm}
\usepackage{color}
\usepackage{verbatim}
\usepackage{cleveref}
\usepackage{url}
\usepackage{balance}

\usepackage{amsfonts}
\usepackage{amsmath}
\usepackage{amssymb}
\usepackage{commath}
\usepackage{cite}
\usepackage{dsfont}
\usepackage{latexsym}
\usepackage{times}
\usepackage{url}
\usepackage{verbatim}

\def\cC{{\mathcal{C}}}

\def\cF{{\mathcal{F}}}

\def\cL{{\mathcal{L}}}

\def\cN{{\mathcal{N}}}

\def\cV{{\mathcal{V}}}

\def\bp{{\mathbf{p}}}

\def\bone{\mathbf{1}}

\def\bA{{\mathbf{A}}}

\def\bE{{\mathbf{E}}}

\def\bH{{\mathbf{H}}}
\def\bI{{\mathbf{I}}}
\def\bJ{{\mathbf{J}}}

\def\bsfH{{\bm{\mathsf{H}}}}

\def\bsfF{{\bm{\mathsf{F}}}}
\def\bsfR{{\bm{\mathsf{R}}}}

\def\bsfx{{\bm{\mathsf{x}}}}
\def\bsfy{{\bm{\mathsf{y}}}}
\def\bsfn{{\bm{\mathsf{n}}}}
\def\bsfs{{\bm{\mathsf{s}}}}
\def\bsff{{\bm{\mathsf{f}}}}
\def\bsfv{{\bm{\mathsf{v}}}}
\def\bsfg{{\bm{\mathsf{g}}}}

\DeclareMathOperator{\blkdiag}{blkdiag}
\DeclareMathOperator{\T}{\sf T}

\def\R{{\mathbb{R}}}
\def\C{{\mathbb{C}}}

\def\Nt{N_{\sf t}}
\def\Ntrf{N_{\sf t}^{\sf RF}}

\def\Nr{N_{\sf r}}

\def\Ns{N_{\sf s}}

\def\cj{{\sf j}}

\def\beeta{{\bm{\eta}}}

\def\e0{{\epsilon_0}}

\def\Nt{N_{\sf t}}
\def\Nr{N_{\sf r}}

\def\Ntrf{N_{\sf t}^{\sf RF}}
\def\Ntuc{N_{\sf t}^{\sf uc}}
\def\Ntpar{N_{\sf t}^{\sf par}}

\def\Ntrad{N_{\sf t}^{\sf rad}}

\def\Ns{N_{\sf s}}

\def\cj{{\sf j}}
\def\dma{{\sf dma}}

\def\ana{{\sf ana}}
\def\hyb{{\sf hyb}}

\def\fro{{\mathsf{fro}}}

\def\Fdig{{\bsfF_{\sf dig}}}
\def\Fana{{\bsfF_{\sf ana}}}
\def\Fhyb{{\bsfF_{\sf hyb}}}
\def\Ftri{{\bsfF_{\sf tri}}}
\def\Fdma{{\bsfF_{\sf dma}}}

\def\Fanac{{\bsfF^*_{\sf ana}}}
\def\Fhybc{{\bsfF^*_{\sf hyb}}}
\def\Fdmac{{\bsfF^*_{\sf dma}}}

\def\Fant{{\bsfF_{\sf ant}}}

\def\Heffhyb{{\widetilde{\bsfH}}_{\hyb}}

\def\Feffhyb{{\widetilde{\bsfF}}_{\hyb}}

\def\trl{\theta_{{\sf{r}},\ell}}
\def\prl{\phi_{{\sf{r}},\ell}}
\def\ttl{\theta_{{\sf{t}},\ell}}
\def\ptl{\phi_{{\sf{t}},\ell}}

\def\at{\mathbf{a}_{\mathsf{t}}}
\def\ar{\mathbf{a}_{\mathsf{r}}}

\def\sfj{{\mathsf{j}}}

\def\dx{d_{\mathsf{x}}}
\def\dy{d_{\mathsf{y}}}
\def\dr{d_{\mathsf{r}}}

\def\Nsub{N_{\mathsf{sub}}}

\def\al{\alpha_{\ell}}
\def\wtau{\omega_{\tau_{\ell}}}
\def\taul{\tau_{{\ell}}}

\def\Lps{L_{\mathsf{PS}}}
\def\Ld{L_{\mathsf{D}}}
\def\Lc{L_{\mathsf{C}}}

\def\ana{\mathsf{ana}}

\def\ant{\mathsf{ant}}

\def\DH{\mathsf{DH}}
\def\TH{\mathsf{TH}}

\def\FD{\mathsf{FD}}
\def\HP{\mathsf{HP}}
\def\HF{\mathsf{HF}}

\newcommand{\edit}[1]{\textcolor{black}{#1}}
\begin{document}

\title{Embracing Reconfigurable Antennas in the Tri-hybrid MIMO Architecture for 6G and Beyond}
\author{Miguel Rodrigo Castellanos,~\IEEEmembership{Member,~IEEE}, Siyun Yang,~\IEEEmembership{Graduate Student Member,~IEEE},\\ Chan-Byoung Chae,~\IEEEmembership{Fellow,~IEEE}, and {Robert W. Heath, Jr.}~\IEEEmembership{Fellow,~IEEE} \\ \vspace{1.5em} \textit{Invited Paper}
\thanks{M. R. Castellanos is with the Department of Electrical Engineering and Computer Science, University of Tennessee, Knoxville, TN 37996, USA. S. Yang and C.-B. Chae$^{*}$ are with the School of Integrated Technology, Yonsei University, Seoul 03722, South Korea. R. W. Heath, Jr.$^{*}$ is with the Department of Electrical and Computer Engineering, University of California, San Diego, La Jolla, CA 92093, USA. *Corresponding authors. This material is based upon work supported by the NSF under grant nos. NSF-ECCS-2435261, NSF-CCF-2435254, NSF ECCS-2414678  and in part by the Army Research Office under Grant W911NF2410107. This work was also supported by IITP under 6G Cloud R\&E Open Hub, and Post-MIMO (IITP-2025-RS-2024-00428780, IITP-2021-0-00486) grants funded by the Korean government.}
}

\markboth{IEEE Transactions on Communications, 2025}%
{Castellanos \MakeLowercase{\textit{et al.}}: Tri-hybrid MIMO for 6G and Beyond}

\maketitle

\begin{abstract}
Multiple-input multiple-output (MIMO) communication has led to immense enhancements in data rates and efficient spectrum management. The evolution of MIMO, though, has been accompanied by increased hardware complexity and array sizes, causing the system power consumption to increase. Despite past advances in power-efficient hybrid architectures, new solutions are needed to enable extremely large-scale MIMO deployments for 6G and beyond. In this paper, we introduce a novel architecture that integrates low-power reconfigurable antennas with both digital and analog precoding. This \emph{tri-hybrid} approach addresses key limitations in traditional and hybrid MIMO systems by improving power consumption and adds a new layer for signal processing. We provide an analysis of the proposed architecture and compare its performance with existing solutions, including fully-digital and hybrid MIMO systems. The results demonstrate significant improvements in energy efficiency, highlighting the potential of the tri-hybrid system to meet the growing demands of future wireless networks. We conclude the paper with a summary of design and implementation challenges, including the need for technological advancements in reconfigurable array hardware and tunable antenna parameters.
\end{abstract}


\section{Introduction}

\IEEEPARstart{M}{IMO} communication technology will remain pivotal for future cellular systems. In the era of 6G and beyond, the demand for massive connectivity, ultra-low latency, and high energy efficiency is driving the exploration of advanced wireless communication technologies \cite{BjoernsonEtAl6GMIMO}. MIMO achieves these goals through spatial multiplexing, high beamforming gains, and multi-user communications. The beauty of MIMO lies in how these benefits all scale with the number of antennas. As MIMO dimensions increase, however, traditional digital architectures--where data can be sent by each antenna-- face challenges related to hardware complexity and power consumption \cite{SkrimponisEtAlTowardsEnergyEfficientMobile2021}. Consequently, communication systems in 6G and beyond will require more efficient transceiver architectures to support a tenfold or greater increase in MIMO dimensions.


Energy efficiency will be a key performance indicator for future wireless networks~\cite{Samsung6GHyperConnectivity2020,5GPPPBeyond5G6GKPI2023,EricssonEmpoweringSustainability2024}. An important factor driving power consumption in MIMO systems is the large number of radio-frequency (RF) chains, which account for both transmit power and the operating power of components in the signal path~\cite{RappaportEtAlWasteFactorAndWaste2024}. Fully digital architectures consume significant power because each antenna requires its own RF chain, including a dedicated data converter. In high-bandwidth millimeter wave (mmWave) systems, the power demand of high-speed data converters becomes a limiting factor when scaling to large arrays. Hybrid architectures mitigate this challenge by reducing the number of RF chains and employing analog precoding techniques, such as phase shifters, true-time delay units, and even lenses, to map a small number of digitally processed signals to a larger array of antennas. Previous research has shown that hybrid architectures can achieve performance comparable to fully digital solutions, particularly at mmWave and sub-THz frequencies
\cite{O.E.AyachEtAlSpatiallySparsePrecodingMillimeter2014,CastellanosEtAlChannelReconstructionBasedHybrid2018, SohrabiYuHybridAnalogAndDigital2017, ParkEtAlDynamicSubarraysHybridPrecoding2017}. Large-scale hybrid MIMO systems, however, will consume enormous amounts of power even with analog precoding and low-resolution data converters \cite{WangTutorialExtremelyLargeScale2024}. In addition, the problem of hardware design becomes more complicated due to the high numbers of antennas and RF components \cite{HanTowardExtraLargeScale2023}. Novel methods are needed to decrease the power consumption of large-scale MIMO without compromising network performance.

\begin{figure*}[!t]
    \centering
    \begin{subfigure}{0.45\textwidth}
        \centering
        \includegraphics[width=\textwidth]{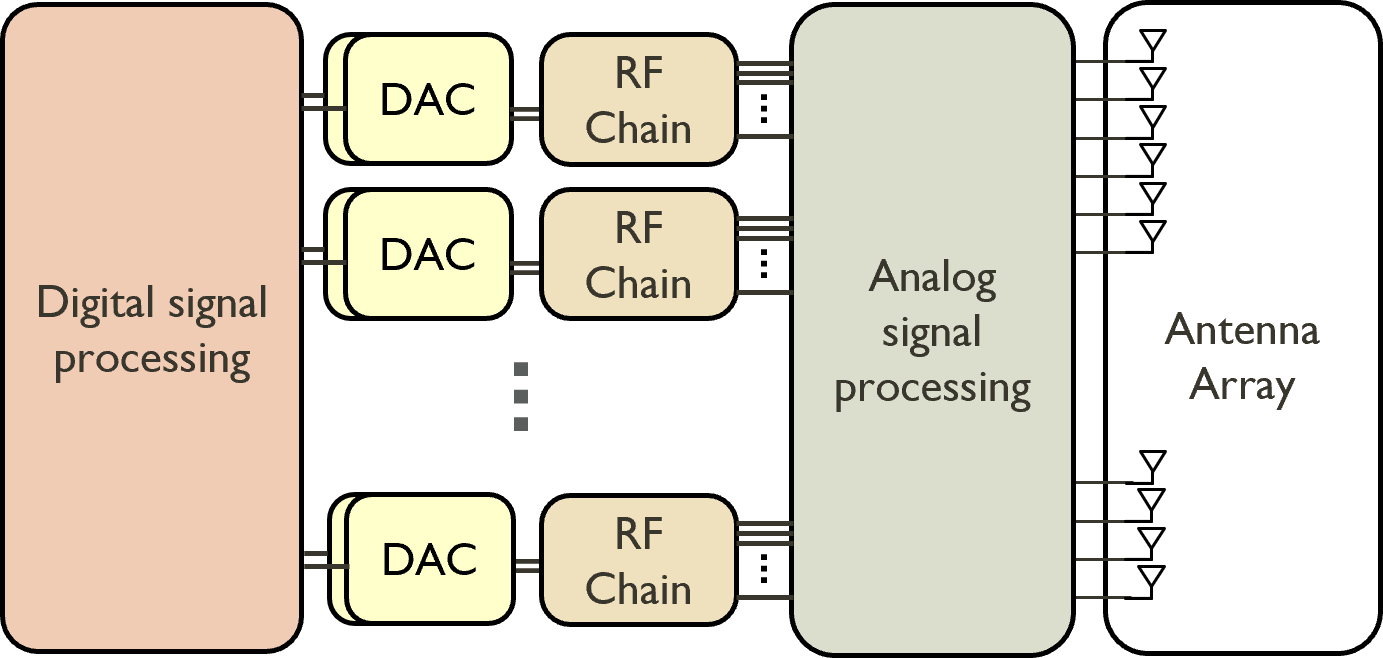}
        \caption{Fully connected hybrid architecture}
        \label{fig:fig1a}
    \end{subfigure}
    \hfill
    \begin{subfigure}{0.45\textwidth}
        \centering
        \includegraphics[width=\textwidth]{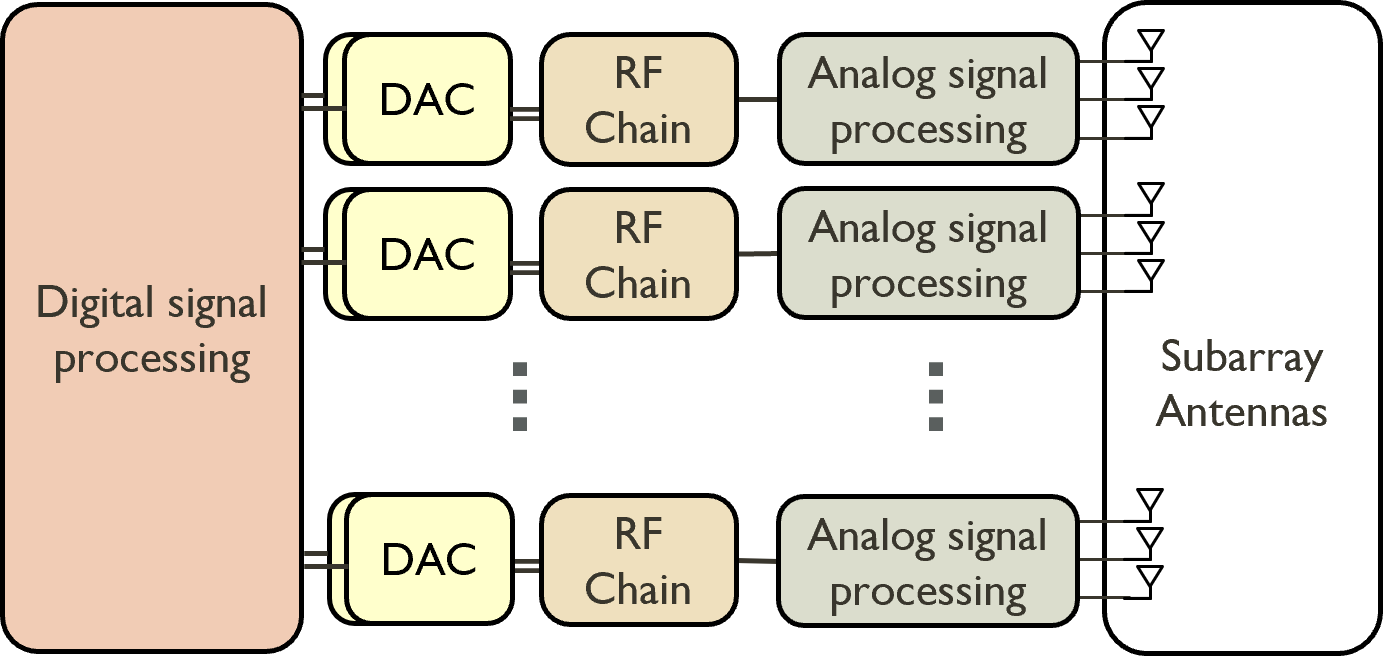}
        \caption{Partially connected hybrid architecture}
        \label{fig:fig1b}
    \end{subfigure}

    \vspace{0.4cm}

    \begin{subfigure}{0.45\textwidth}
        \centering
        \includegraphics[width=\textwidth]{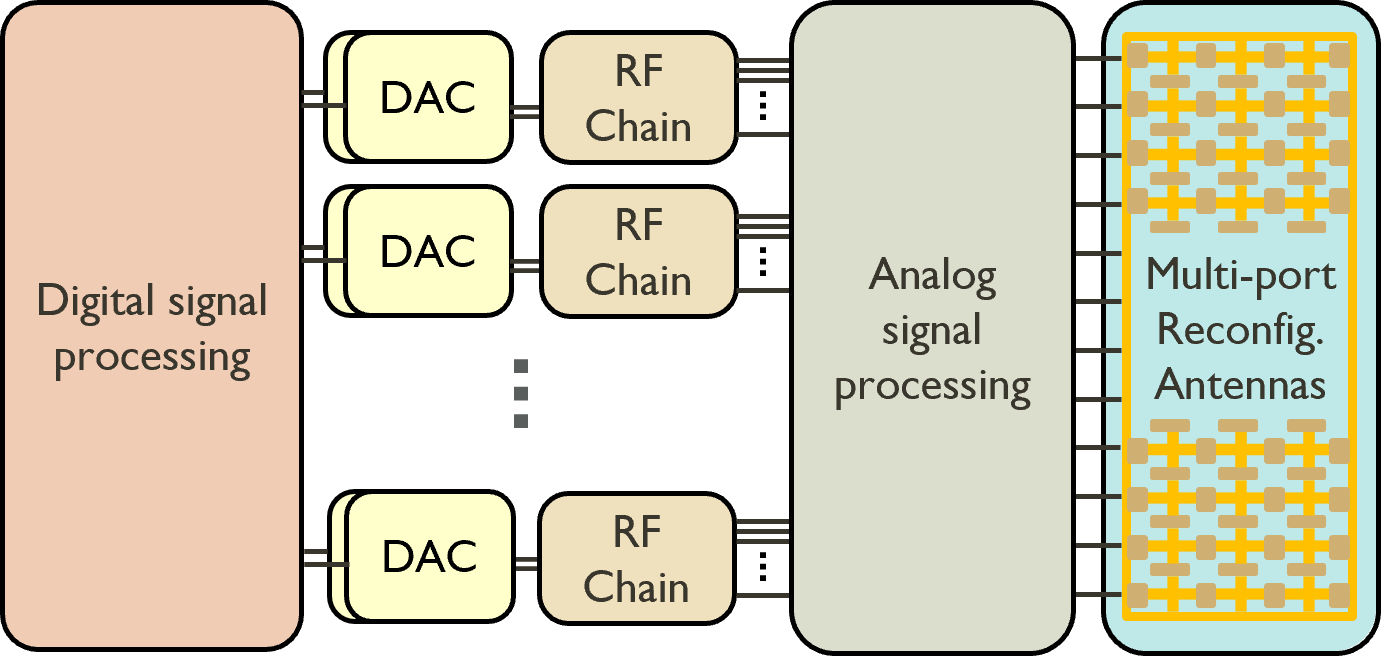}
        \caption{Fully connected tri-hybrid architecture}
        \label{fig:fig1c}
    \end{subfigure}
    \hfill
    \begin{subfigure}{0.45\textwidth}
        \centering
        \includegraphics[width=\textwidth]{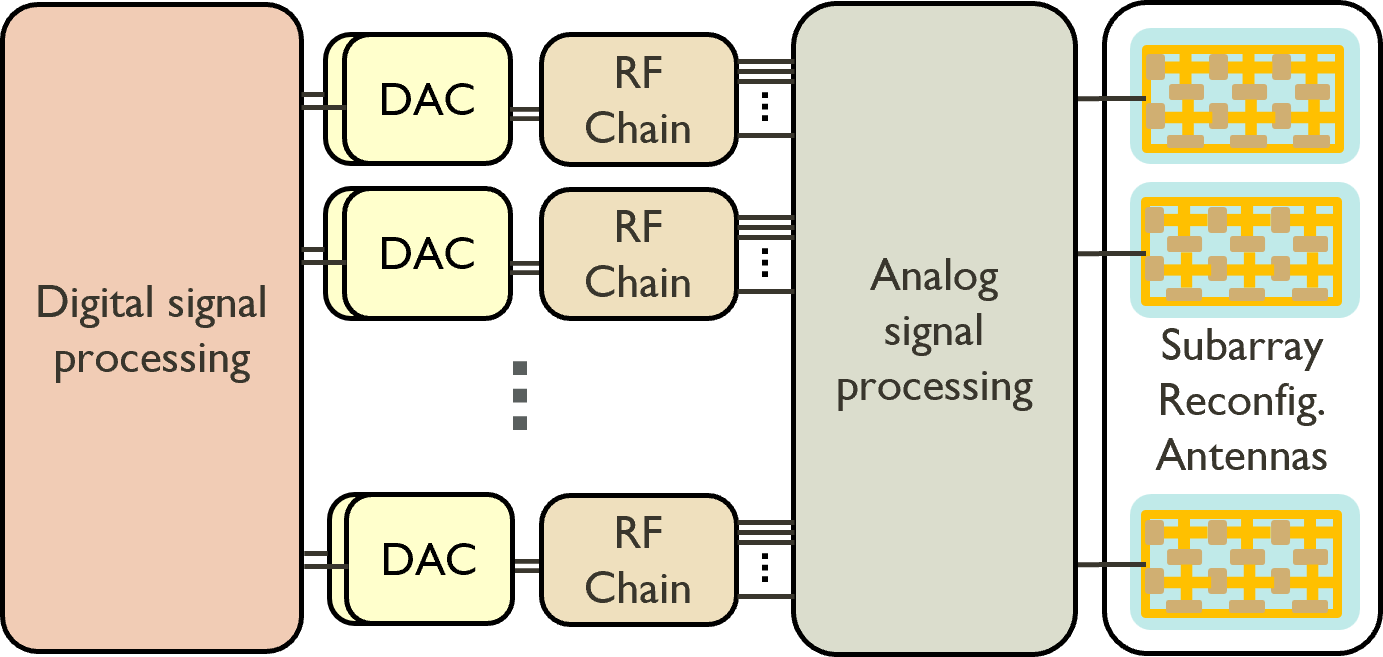}
        \caption{Partially connected tri-hybrid architecture} 
        \label{fig:fig1d}
    \end{subfigure}
    \caption{Illustration of different MIMO system architectures. (a) Fully-connected hybrid (5G NR), (b) partially-connected hybrid (5G NR), (c) fully-connected multi-port tri-hybrid (proposed for 6G and beyond), (d) fully connected reconfigurable antenna for tri-hybrid (proposed for 6G and beyond). The partially connected analog structure can be combined with multi-port reconfigurable metasurfaces like in (c) or with reconfigurable antennas as shown in (d).}  
    \label{fig:mimo_architectures}
\end{figure*}

Reconfigurable arrays are paving the way for novel MIMO architectures capable of low-power spatial signal processing. \edit{Reconfigurable antennas are dynamic devices that can change their operating state \cite{HauLan:Reconfigurable-Antennas:13}.} Different configurations can correspond to distinct polarizations, gain patterns, frequency responses, or a combination of any number of antenna characteristics. Reconfiguration methods are varied, with designs leveraging electromechanical switches \cite{AnaGuiChr:RF-MEMS-Reconfigurable:06}, metasurfaces \cite{LiChe:Dual-Band-Metasurface:18}, and conductive fluids \cite{BhaMaDic:RESHAPE:-A-Liquid-Metal-Based:21,chae_fluid1,chae_fluid2}. Most reconfigurable antennas share the common feature of relatively low power consumption compared to \edit{a similarly sized aperture with analog signal processing.} Reconfigurable antennas, therefore, offer a promising avenue for energy-efficient MIMO systems that also \edit{extend signal processing to antennas.}

In this paper, we propose the novel \emph{tri-hybrid} architecture that combines hybrid precoding with highly-reconfigurable antenna arrays. This design leverages the flexibility of reconfigurable antennas to enhance system performance while maintaining energy efficiency. Hybrid architectures, as deployed in 5G New Radio (5G NR), divide beamforming operations between digital and analog domains, thereby facilitating the implementation of larger antenna arrays. For instance, Fig.~\ref{fig:mimo_architectures}(a) illustrates a fully connected hybrid solution, while Fig.~\ref{fig:mimo_architectures}(b) uses a partially connected hybrid scheme. The tri-hybrid architecture replaces static antenna arrays with reconfigurable apertures. This architecture encompasses various kinds of reconfigurable array designs in literature. Fig.\ref{fig:mimo_architectures}(c) shows how the entire array can be replaced with a multi-port reconfigurable surface, while Fig.\ref{fig:mimo_architectures}(d) substitutes each array element with a subarray of reconfigurable antennas. 

We first describe the reconfigurable antenna functionality from an electromagnetic perspective. In Section \ref{sec_recon}, we provide an overview of how surface current distributions determine antenna characteristics and show how changes in antenna currents lead to reconfigurable behaviors. We also discuss how each kind of reconfigurable antenna can be applied in the context of wireless communications. This broad overview of reconfigurable array functionality is critical for modeling tri-hybrid MIMO systems.

We then provide an introduction to various tri-hybrid MIMO architectures, highlighting their structural designs, functionalities, and applications in Section \ref{sec_survey}. We explore the different configurations and integration methods used to combine distinct technologies or systems into a cohesive architecture. Additionally, we discuss the selection criteria for reconfigurable antennas within these architectures, examining factors such as adaptability, performance metrics, and suitability for specific applications. By surveying the existing types and evaluating the options for reconfigurable antennas, \edit{the section provides a comprehensive outlook of the current landscape and future developments in tri-hybrid system designs.}

We then discuss analytical models of a tri-hybrid architecture that integrate reconfigurable antennas with both digital and analog precoding in Section \ref{sec_tri_hybrid}. \edit{Effectively, the reconfigurable antenna becomes a complementary layer to perform electromagnetic (EM) precoding~\cite{dardari2024DSA}.} In Section \ref{sec_performance}, we analyze the tri-hybrid architecture and the impact of the reconfigurable array on the system model \edit{in terms of both spectral efficiency and power consumption}. Subsequently, we examine a specific architecture using a particular type of reconfigurable antenna known as a dynamic metasurface antenna (DMA). We develop a baseline algorithm for tri-hybrid precoding and compare it with traditional MIMO solutions in terms of spectral efficiency and power consumption. Our findings demonstrate the potential of the tri-hybrid approach to meet the demands of next-generation wireless networks.

\section{What are reconfigurable antennas?}
\label{sec_recon}

Before introducing the tri-hybrid architecture, we discuss the use of reconfigurable arrays in the context of wireless communication. We approach reconfigurability from an electromagnetic perspective to form a physically-consistent foundation for modeling and analyzing tri-hybrid systems. We also discuss practical methods for antenna reconfiguration.

\subsection{Fundamentals of reconfigurability}

A reconfigurable antenna can be dynamically controlled to adjust its properties such as polarization, gain, and frequency. The radiative characteristics of an antenna are largely determined by its physical properties. Environmental factors, such as temperature, humidity, and even nearby objects can also affect how antennas operate. In contrast with a static antenna, a reconfigurable antenna is able to adapt how it radiates through intentional and reversible operations. A simple design, for example, could leverage PIN diodes or other types of electronic switches to redirect the antenna feed between two differently sized or oriented antennas. More sophisticated designs leverage a combination of tunable metamaterials and electromechanical devices to enable several operating modes.

Reconfigurable antennas change states by altering the antenna current distribution and, therefore, the radiated electromagnetic fields. The antenna surface currents determine how it behaves and radiates. Let $\cV \subseteq \R^3$ denote the volume that encompasses the antenna. At any point $\bp \in \cV$, we let $\bJ(\bp)$ be the current density in the antenna. The radiated electric and magnetic fields of an antenna can be computed using an auxiliary quantity known as the magnetic vector potential, denoted as $\bA(\bp)$. Let $c$ be the speed of light, $f$ be the operating frequency, $\lambda = c/f$ be the operating wavelength and $\beta = 2\pi / \lambda$ be the angular wavenumber. We also define
$\varepsilon_0$ as the vacuum permittivity and $\mu_0$ as the vacuum permeability. The magnetic vector potential induced by the antenna current distribution is
\begin{equation}
\label{eq_vector_potential}
    \bA(\bp) = \mu_0 \int_\cV \frac{e^{\cj \beta \norm{\bp - \bp'}}}{4 \pi \norm{\bp - \bp'}} \bJ(\bp') \, d \bp'.
\end{equation}
The vector $\bA(\bp)$ is used to compute the induced electromagnetic fields at any point $\bp$ in space through the following expressions:
\begin{equation}
\label{eq_magnetic_field}
    \bH(\bp) = \frac{1}{\mu_0} \nabla \times \bA(\bp),
\end{equation}
\begin{equation}
\label{eq_electric_field}
    \bE(\bp) = \frac{1}{\cj \omega \epsilon_0} \nabla \times \bH(\bp).
\end{equation}
The field expressions in \eqref{eq_vector_potential}, \eqref{eq_magnetic_field} and \eqref{eq_electric_field} demonstrate that the antenna properties are fully dependent on the current distribution $\bJ(\bp)$ and the volume $\cV$. In the following, we overview how these quantities are manipulated to enable antenna reconfiguration.

\subsection{Polarization}
An antenna can change its polarization by reorienting the surface current. The polarization of an antenna can be defined as the orientation of the electric field that makes up the radiated wavefronts. For example, a horizontally polarized antenna will radiate an electric field that is aligned with the azimuthal plane. The magnetic field and the electric field orientations are connected to $\bA(\bp)$ through \eqref{eq_magnetic_field} and \eqref{eq_electric_field}. Let $\bA(\bp)$ be written in terms of radial component $\bA_{\sf rad}(\bp)$, its azimuthal component $\bA_{\sf azi}(\bp)$ and its elevation component $\bA_{\sf ele}(\bp)$. In the far-field ($\norm{\bp} \rightarrow \infty$), the nonradial electric field components become approximately colinear with the magnetic vector potential as
\begin{equation}
    \bE(\bp) \approx - \cj \omega \left[0,\, \bA_{\sf azi}(\bp), \, \bA_{\sf ele}(\bp) \right]^T.
\end{equation}
From \eqref{eq_vector_potential}, we see that $\bA(\bp)$ is largely aligned with the \edit{dominant orientation of the surface current}. A simple example of this is a dipole antenna: the electric current, and therefore the radiated electric field, is oriented in the same direction as the dipole axis. Changing the antenna surface current direction is the key mechanism that enables polarization reconfiguration. The number of achievable current directions on the antenna then determines the number of polarization states.

Reconfigurable polarization arrays can effectively combat transceiver misalignments and promote polarization diversity \cite{ZhuEtAlDesignPolarizationReconfigurableAntenna2014}. Polarization mismatches between antennas and impinging wavefronts cause losses in receive power. Polarization reconfiguration remedies this issue by adaptively changing the antenna polarization without significantly affecting the gain pattern \edit{or needing twice as many elements}. Common polarization modes of operation include linear polarization such as vertical, horizontal, and $\pm45^\circ$ slant, as well as right-handed and left-handed circular polarizations \cite{KhidreEtAlCircularPolarizationReconfigurableWideband2013, GrauEtAlDualLinearlyPolarizedMems2010, LinWongWidebandCircularPolarizationReconfigurable2015}. Prior work demonstrates that polarization reconfiguration is advantageous for both line-of-sight cases where orientation affects polarization, and for non-line-of-sight scenarios with channel depolarization \cite{GrauEtAlDualLinearlyPolarizedMems2010}.

\subsection{Gain pattern}
\edit{Gain patterns are reconfigured} by changing the antenna current distribution between different patterns. An antenna's gain pattern describes \edit{the strength of radiation and absorption as a function of direction}. The relationship between the antenna gain pattern and $\bJ(\bp)$ is not as intuitive as that between the surface current orientation and the polarization, but it can still be exploited. One approach is to simply mimic the current distribution of different antenna types with known antenna patterns \cite{ChengAzimuth2017}. Similar to array pattern synthesis, desired antenna patterns can be inversely mapped to current distributions, which can then guide the desired current configurations~\cite{SalucciSynthesis2018}. While generally more challenging than polarization reconfiguration, gain pattern reconfiguration effectively enables antennas to \edit{individually beamform}.

\begin{figure*}[!t]
    \centering
    \begin{minipage}[b]{0.2\textwidth}
        \centering
        \includegraphics[width=\textwidth]{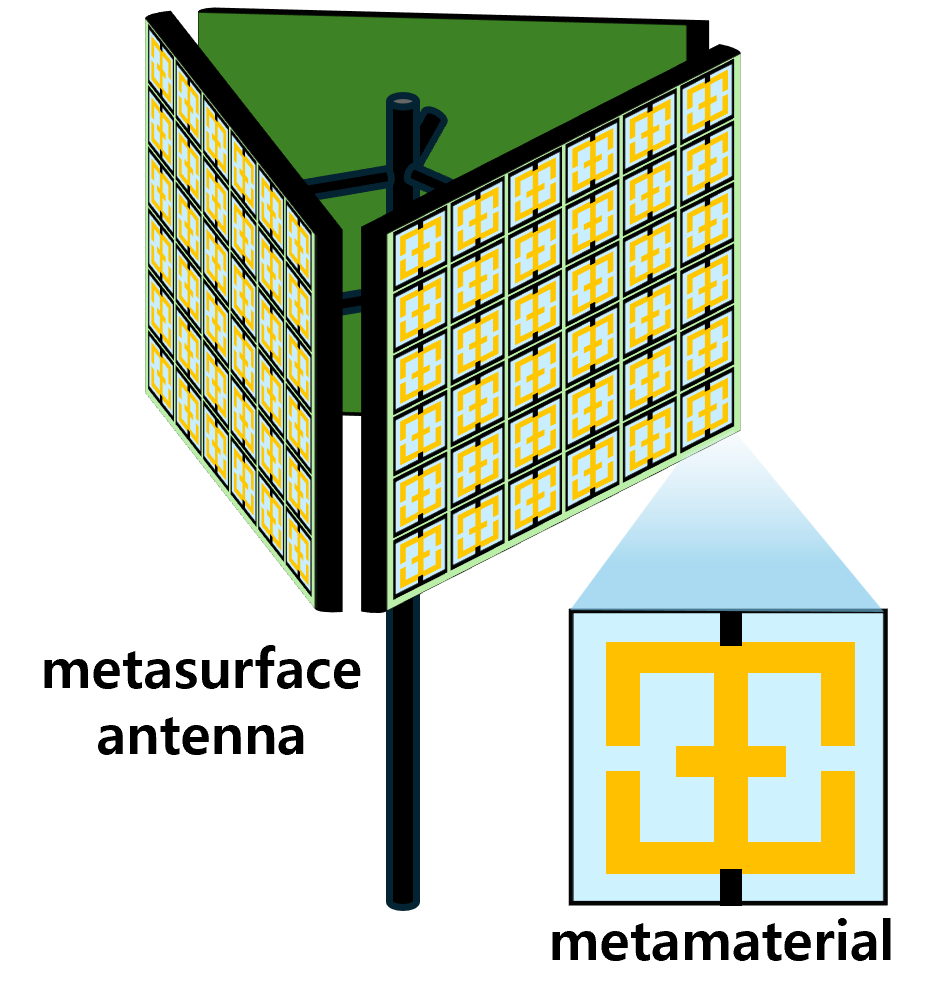}
        \subcaption{Metasurface} 
        \label{fig:sub1}
    \end{minipage}
    \hspace{0.05\textwidth} 
    \begin{minipage}[b]{0.25\textwidth}
        \centering
        \includegraphics[width=\textwidth]{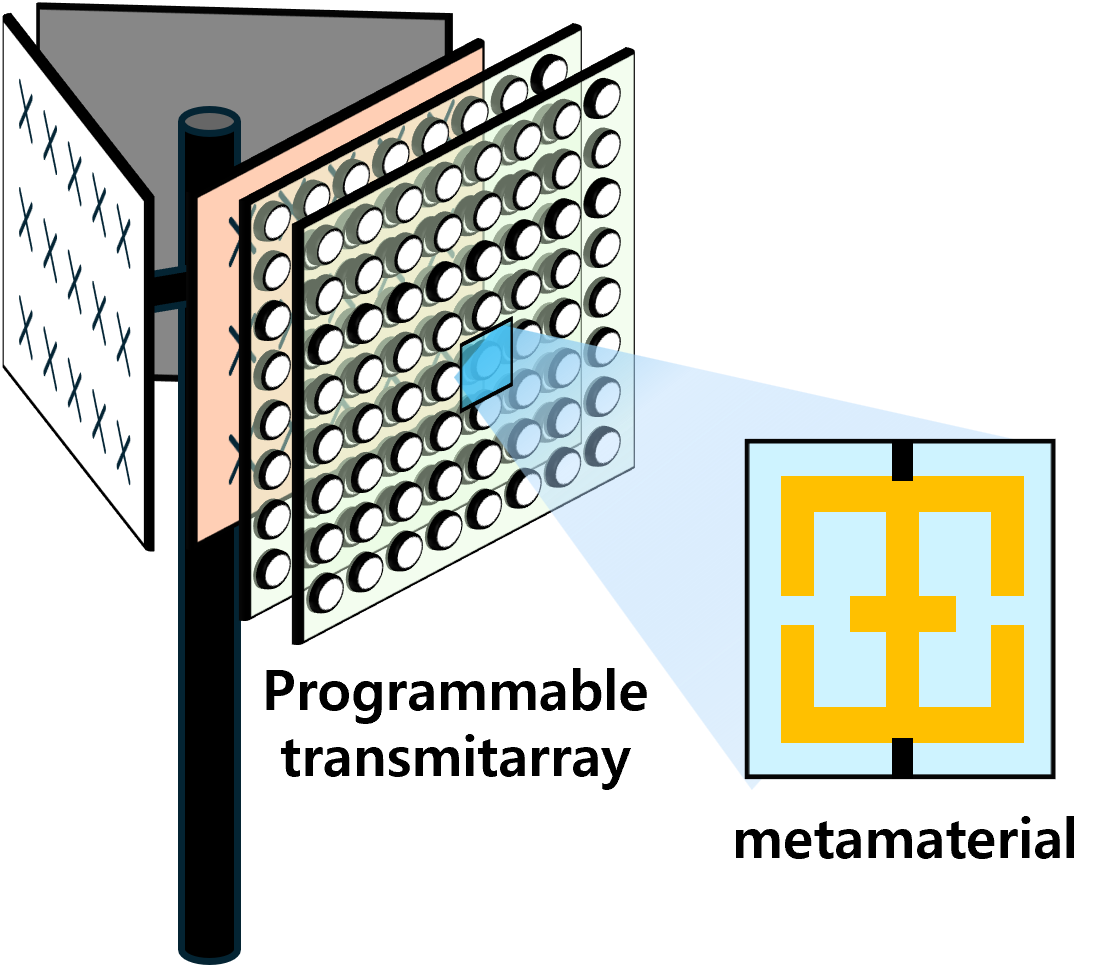}
        \subcaption{Parasitic element-assisted} 
        \label{fig:sub1}
    \end{minipage}
    \hspace{0.05\textwidth} 
    \begin{minipage}[b]{0.3\textwidth}
        \centering
        \includegraphics[width=\textwidth]{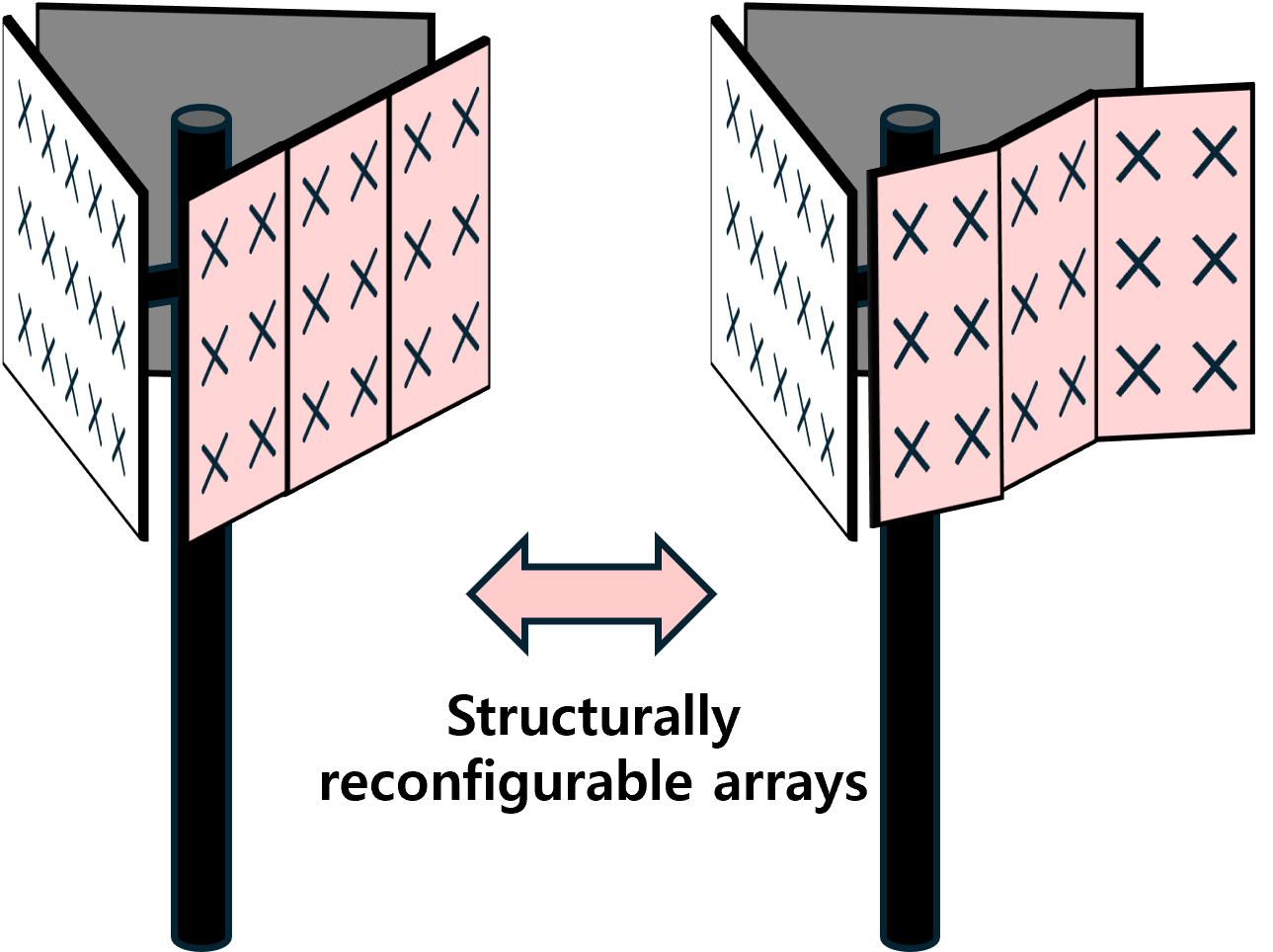}
        \subcaption{Structurally reconfigurable} 
        \label{fig:sub1}
    \end{minipage}

    \caption{\edit{Examples of reconfigurable antenna designs that enable dynamic control of radiation patterns and operating frequencies. 
    (a) \textit{Metasurface approach} leverages reprogrammable or tunable surface elements to manipulate electromagnetic waves and achieve beam steering. 
    (b) \textit{Parasitic element-assisted approach} adjusts the arrangement or state of active or passive parasitic elements to alter the antenna’s effective aperture and radiation characteristics. 
    (c) \textit{Structurally reconfigurable approach} physically modifies the geometry or mechanical configuration of the antenna, thereby providing flexible beam shaping and coverage.}}

    \label{fig:Various type of reconfigurable antennas}
\end{figure*}

Gain pattern reconfiguration enables individual antennas to beamform without the need for external signal processing components \cite{HuffBernhardIntegrationPackagedRfMems2006}. Different use-cases necessitate different kinds of operating states. An omnidirectional antenna exhibits a completely symmetrical pattern in space and is useful for rich scattering environments and for initial access. A directional antenna is biased towards particular angles, which mitigates interference and is well-suited for line-of-sight \edit{channels}. Reconfigurable antennas are beneficial in dynamic environments with \edit{high mobility or frequent blockages} \cite{ZhangEtAlPatternReconfigurableMicrostripParasitic2004,QinEtAlPatternReconfigurableUSlot2012}. \edit{Pattern null reconfiguration reduces interference to other devices in the network, which is beneficial for spectrum co-existence at saturated frequencies such as the upper mid-band \cite{KangWireless2024}.} While the array pattern can also be adapted through digital and RF processing components, antenna reconfiguration leads to overall lower power consumption and hardware complexity.

\subsection{Frequency response}
The frequency response of an antenna can also be reconfigured by emulating a desired current distribution at different operating frequencies. Thus far, we have neglected the dependence of the current distribution on the frequency $f$. In reality, $\bJ(\bp, f)$ is a function of both space and frequency. 
One goal of frequency reconfiguration is to alter the antenna such that the same current distribution is achievable at distinct frequencies, thus enabling multiple operating frequencies. The easiest way to achieve this is by changing the antenna geometry to adjust the resonance. For example, a reconfigurable half-wave dipole could leverage electronic switches to switch between two lengths, with each length corresponding to a different frequency. Frequency reconfiguration, however, encompasses a broad set of behaviors that change antenna frequency-selectivity  for multi-band narrowband operation, tunable antenna bandwidths, and even frequency rejection for interference management. 
Narrowband antennas, for example, can be reconfigured for wideband operation by generating a current distribution $\bJ(\bp, f)$ with a larger support in terms~$f$. Frequency tunability can greatly simplify system design by enabling multi-band operation without the need for separate antennas for each band. 

In this context, frequency reconfiguration has been leveraged for wideband and multi-band operation 
In the simplest case, reconfigurable narrowband antennas can dynamically tune their operating frequency without changing other antenna properties \cite{PeroulisEtAlDesignReconfigurableSlotAntennas2005,ErdilEtAlFrequencyTunableMicrostripPatch2007,CetinerEtAlRfMemsIntegratedFrequency2010}. More advanced designs have demonstrated the ability to switch between wideband and narrowband \cite{QinEtAlWidebandToNarrowbandTunable2015} and to introduce tunable rejection notches \cite{Perruisseau-CarrierEtAlModelingDesignAndCharacterization2010, NikolaouEtAlUWBEllipticalMonopolesWith2009}. These functions can be exploited to operate in multiple bands with the same front-end and to reduce interference transmission and reception \cite{KumarFifthGenerationAntennas2020}.

\section{Survey of state-of-the-art reconfigurable antennas for tri-hybrid architecture designs }
\label{sec_survey}

Reconfigurable antennas have evolved to feature smaller form factors and offer a greater variety of configurations over the past few decades~\cite{alu}. In this section, we investigate the latest reconfigurable antenna designs to identify potential candidates for integration into a tri-hybrid MIMO architecture. We specifically focus on three main categories, as illustrated in Fig.~\ref{fig:Various type of reconfigurable antennas}: metasurface-based antennas, parasitic element-assisted antennas, and structurally reconfigurable antennas. These advanced designs offer greater flexibility in terms of frequency tuning, radiation pattern control, and polarization manipulation. By exploring the key advancements, design principles, and potential applications of each category, we evaluate their suitability for integration into the tri-hybrid architecture and highlight their role in shaping the future of wireless networks. We acknowledge that other types of reconfigurable antennas can be incorporated into the tri-hybrid architecture, which is a ripe area for future research.

\subsection{Metasurface antennas}

\begin{figure*}[!t]
    \centering
    \begin{minipage}[b]{0.6\textwidth}
        \centering
        \begin{minipage}[b]{0.45\textwidth}
            \centering
            \includegraphics[width=\textwidth]{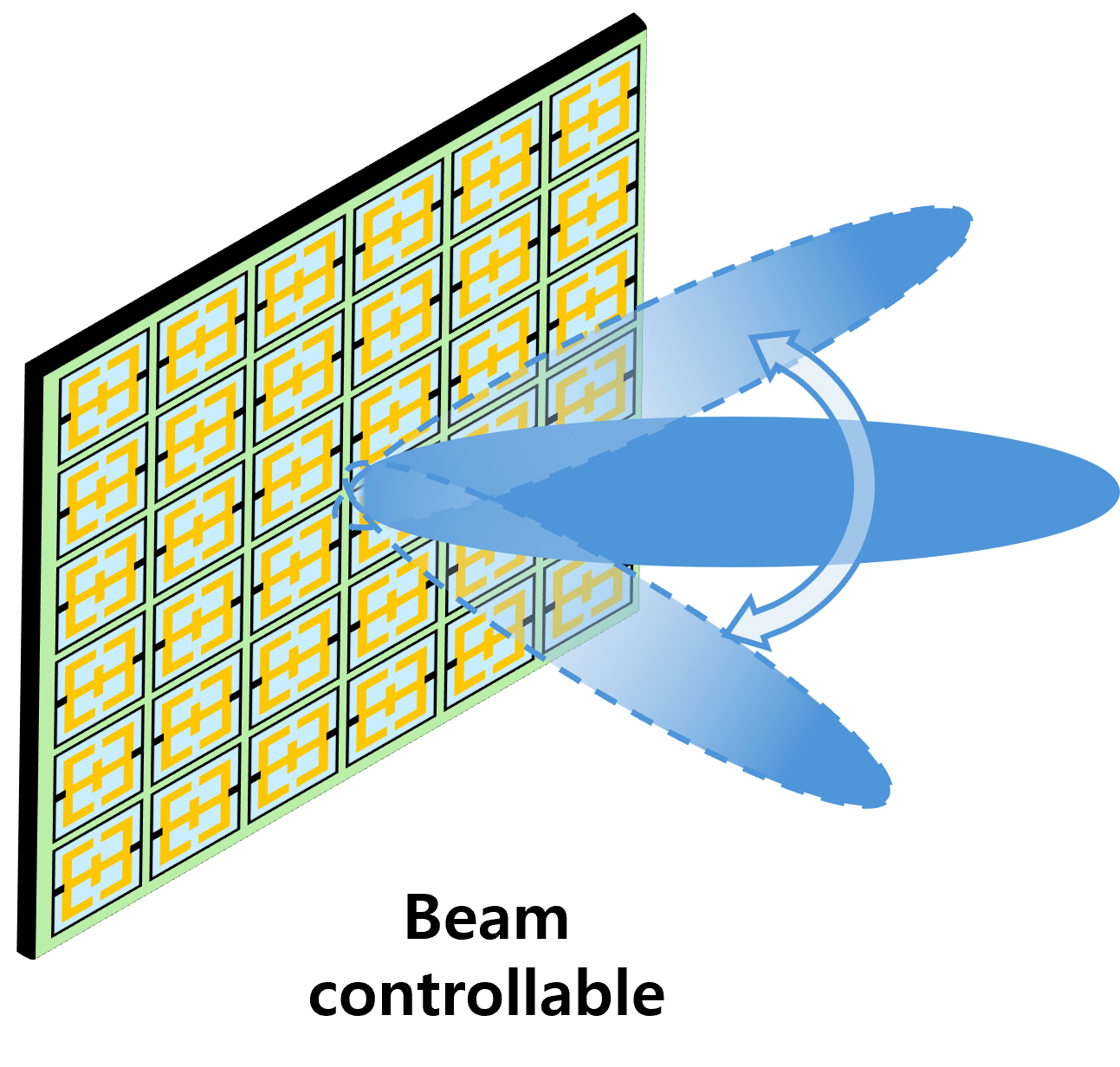}
            \subcaption{Beamforming control}
            \label{fig:DMA Beamforming control}
        \end{minipage}
        \hfill
        \begin{minipage}[b]{0.45\textwidth}
            \centering
            \includegraphics[width=\textwidth]{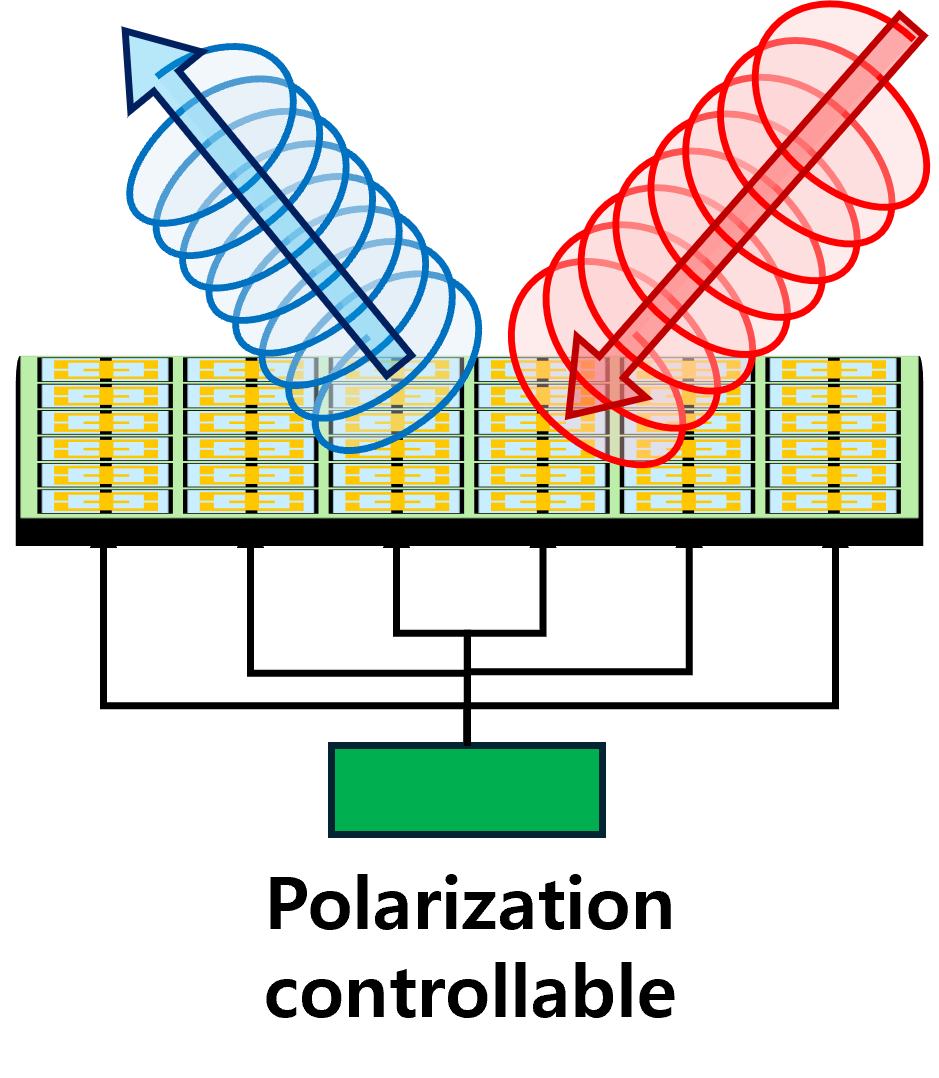}
            \subcaption{Polarization control}
            \label{fig:DMA Polarization control}
        \end{minipage}
        \vskip\baselineskip
        \begin{minipage}[b]{\textwidth}
            \centering
            \includegraphics[width=\textwidth]{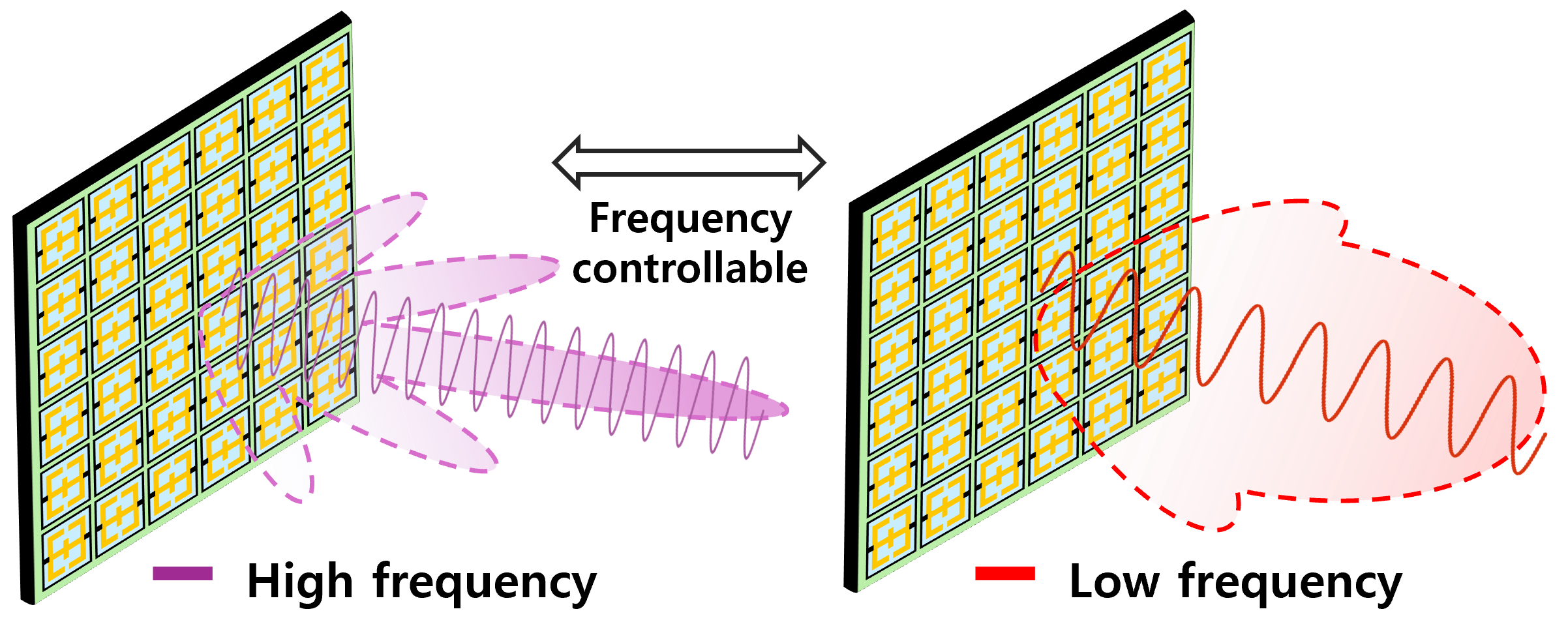}
            \subcaption{Frequency resonance control}
            \label{fig:DMA Frequency resonance control}
        \end{minipage}
    \end{minipage}
    \caption{Key functionalities of metasurface antennas:
    (a) Beamforming control for dynamic steering via phase/amplitude tuning,
    (b) Polarization control for diverse electric field orientations,
    (c) Frequency resonance control to optimize coverage or spectral efficiency.} 
    \label{fig:Various function of metasurface antennas}
\end{figure*}

Metasurface antennas are a highly suitable choice for the antenna layer in a tri-hybrid architecture due to their multifunctionality, adaptability, and proven efficiency. As shown in Fig. \ref{fig:Various function of metasurface antennas}, metasurface antennas can perform various functions depending on their design. These antennas use periodic sub-wavelength resonators, enabling low-profile designs, wide bandwidths, and rapid response times. Previous research has demonstrated that metasurfaces can reduce mutual coupling, enhance antenna gain, and \edit{reduce} power consumption~\cite{li2018metasurfaces, gupta2017mutual, saad2019printed, jiang2019symmetrical, iqbal2019electromagnetic, murthy2020improved}. 

\begin{figure*}[!t]
    \centering
    \begin{minipage}[a]{1.0\textwidth}
        \centering
        \includegraphics[width=0.8\textwidth]{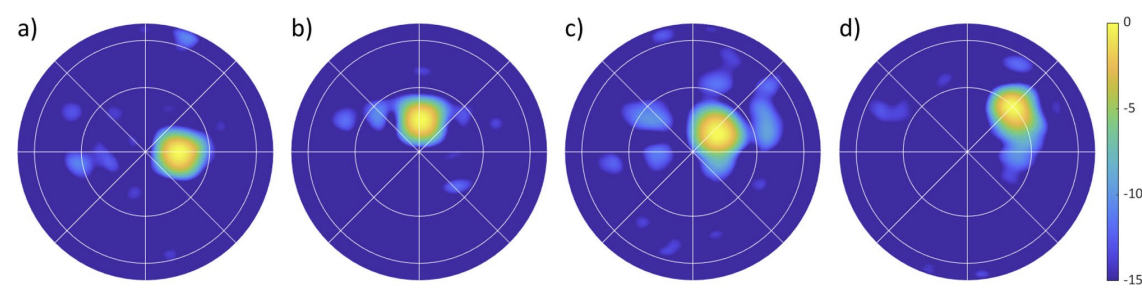}
        \subcaption{Beamforming simulation results for various angles from \cite{Boyarsky2021}}
        \label{fig:sub1}
    \end{minipage}
    \hfill
    \begin{minipage}[b]{0.9\textwidth}
        \centering
        \includegraphics[width=0.9\textwidth]{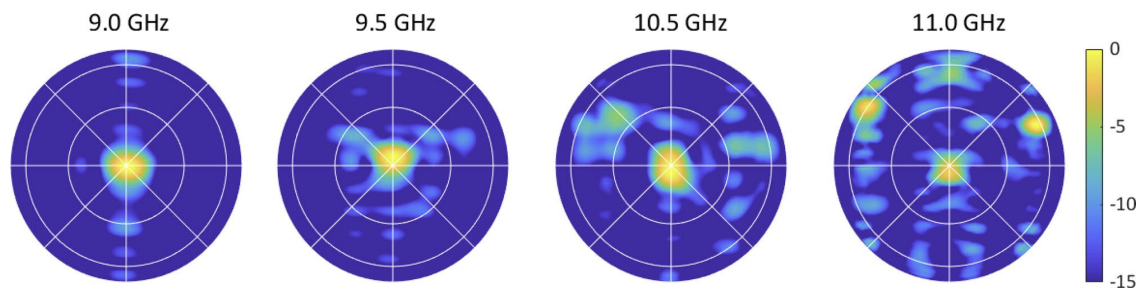}
        \subcaption{Beamforming simulation results for various frequency band from~\cite{Boyarsky2021}}
        \label{fig:example}
    \end{minipage}
    \caption{Simulation results demonstrating the Nyquist metasurface antenna's performance. (a) Depicts the antenna's beam steering capabilities in azimuth and elevation through feed phase diversity, achieving precise and efficient control. (b) Shows beamforming performance across various frequency bands, with the antenna maintaining stable operation within the range of 9.0–11.0 GHz, as detailed in \cite{Boyarsky2021}. }
    \label{fig:DMA based tri-hybird simulation results}
\end{figure*}

Dynamic metasurface antennas (DMAs), \edit{which are leaky waveguides with tunable sub-wavelength slots}, have been successfully used for beamforming across multiple frequency bands~\cite{Boyarsky2021}. By utilizing low-power components such as varactors and PIN diodes, metasurface antennas facilitate efficient reconfiguration. For instance, Fig. \ref{fig:DMA based tri-hybird simulation results} demonstrates the beamforming gain of a DMA as implemented in~\cite{Boyarsky2021}, underscoring the potential of DMAs within tri-hybrid architectures where low-power adaptive beamforming across multiple frequency bands is critical.

\begin{figure*}[!t]
    \centering
    \begin{subfigure}[b]{0.4\textwidth}
        \centering
        \includegraphics[width=0.7\textwidth]{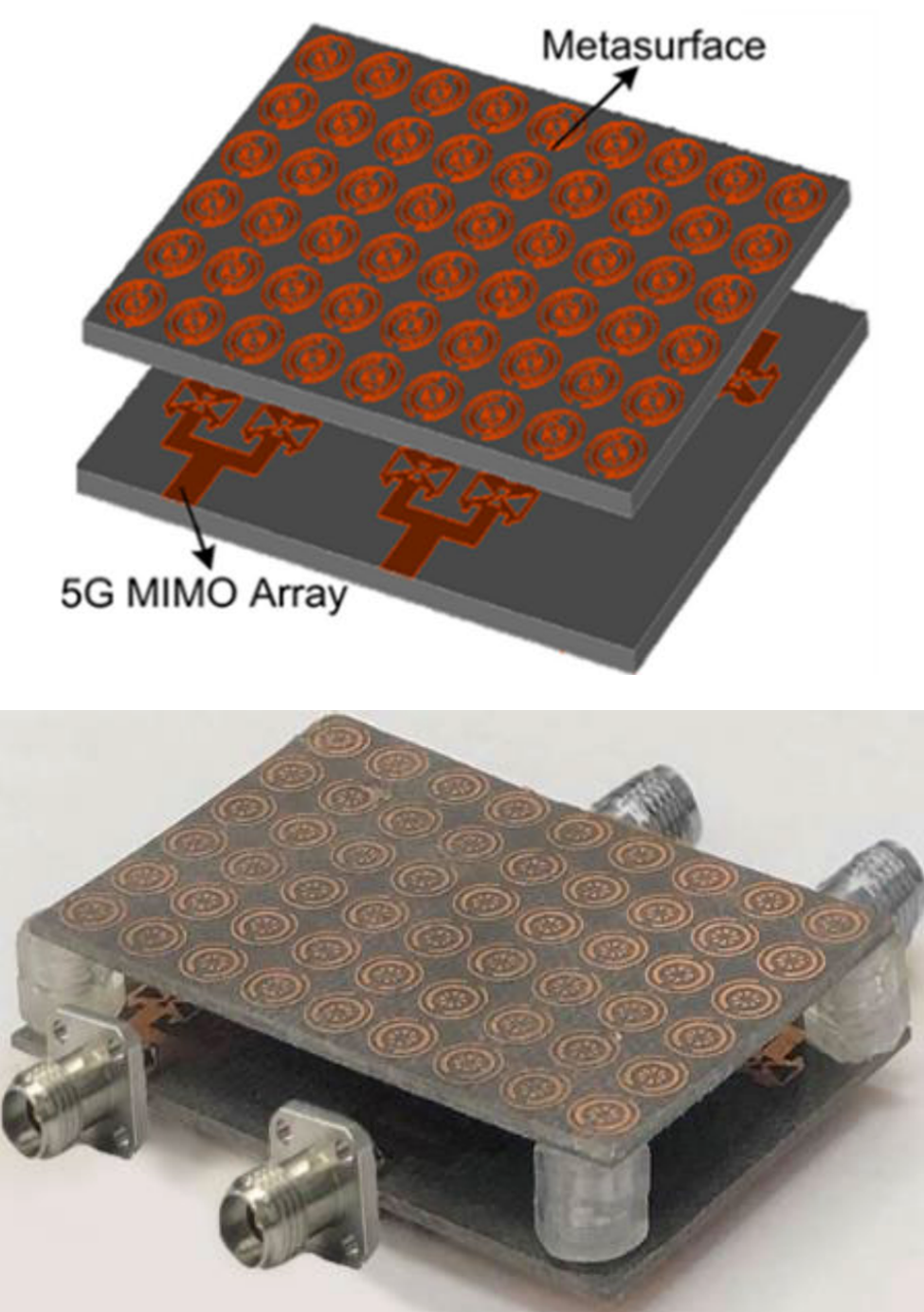}
        \caption{Metasurface with MIMO array \cite{tariq2021metasurface}}
        \label{fig:subfig1}
    \end{subfigure}
    \hspace{0.5cm}
    \begin{subfigure}[b]{0.4\textwidth}
        \centering
        \includegraphics[width=0.7\textwidth]{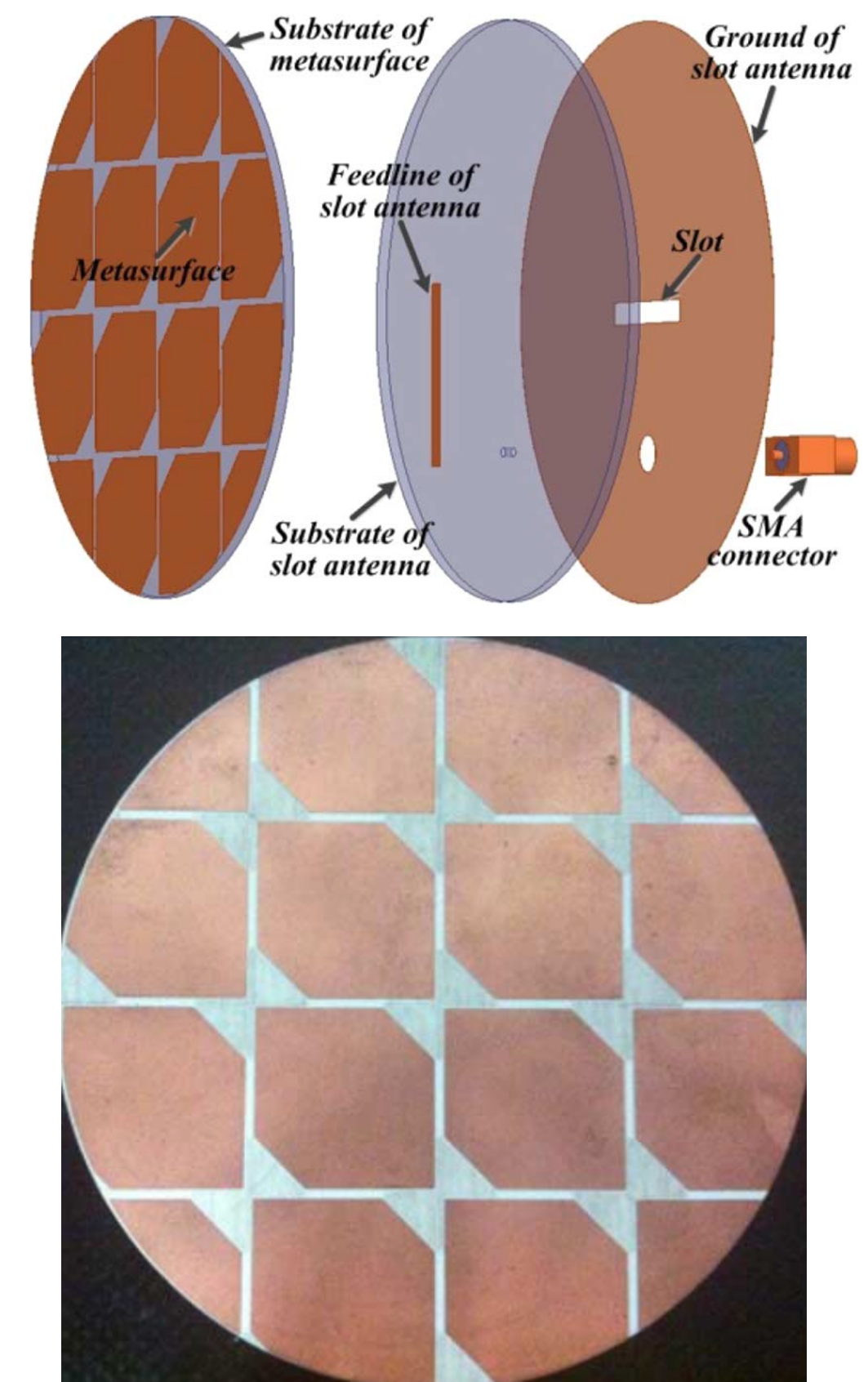}
        \caption{Slot antenna with metasurface \cite{zhu2014design}}
        \label{fig:subfig2}
    \end{subfigure}
    
    \vspace{0.5cm} 

    \begin{subfigure}[b]{0.4\textwidth}
        \centering
        \includegraphics[width=0.7\textwidth]{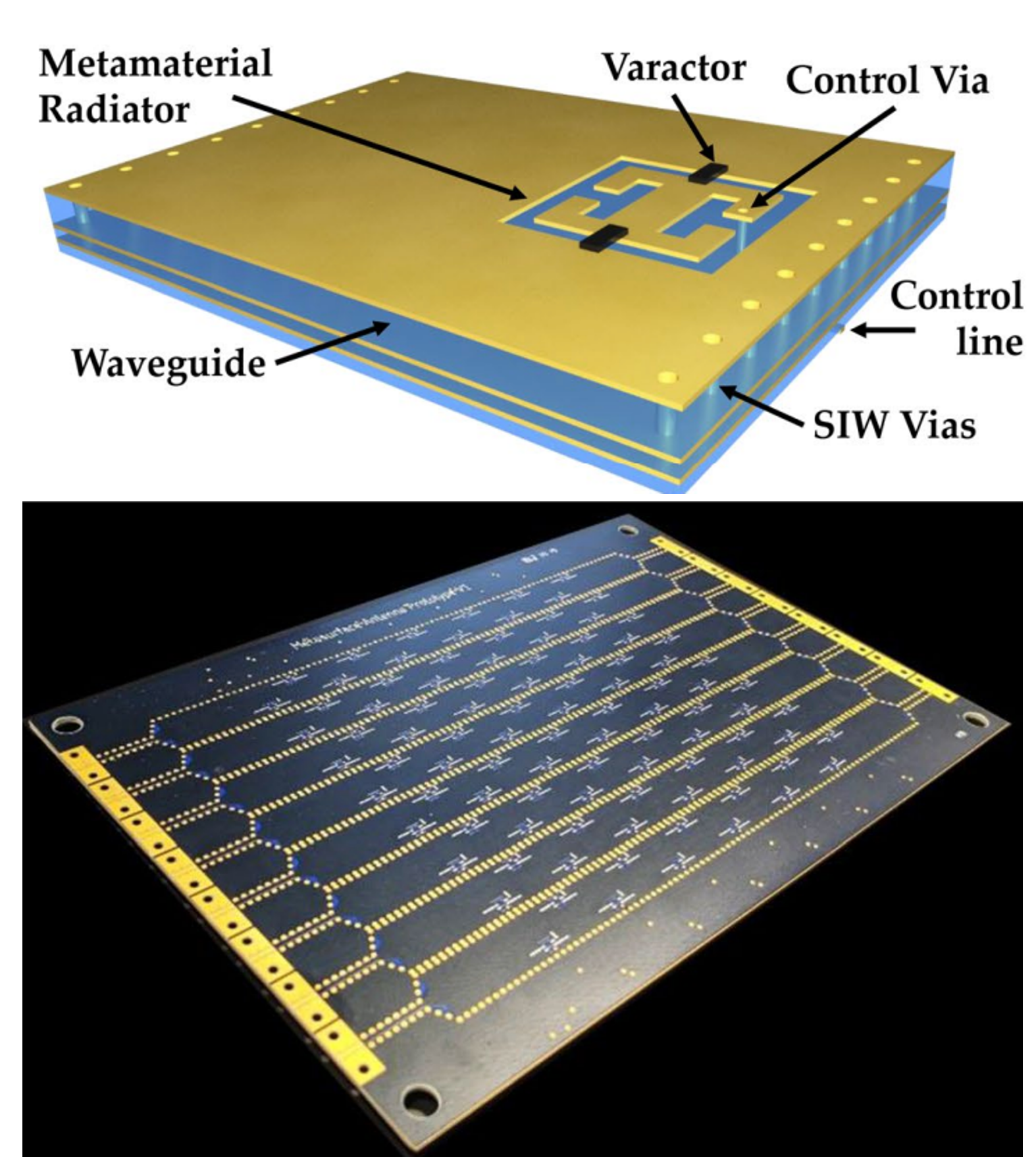}
        \caption{Varactor controlled metasurface \cite{Boyarsky2021}}
        \label{fig:subfig3}
    \end{subfigure}
    \hspace{0.5cm}
    \begin{subfigure}[b]{0.4\textwidth}
        \centering
        \includegraphics[width=0.7\textwidth]{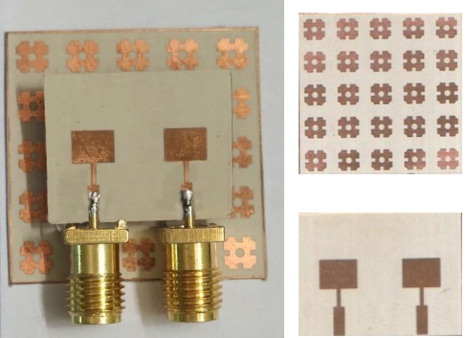}
        \caption{Metasurface unit cells \cite{iqbal2019electromagnetic}}
        \label{fig:subfig4}
    \end{subfigure}
    
    \caption{Various configurations and applications of EM precoding in antenna array design.
    (a) A metasurface reconfigures spatial coupling to improve gain and reduce mutual coupling \cite{tariq2021metasurface}. 
    (b) A polarization-reconfigurable slot antenna adjusts its polarization state by rotating the metasurface \cite{zhu2014design}. 
    (c) A varactor-controlled metasurface dynamically tunes phase distribution for electronic beam steering \cite{Boyarsky2021}. 
    (d) An electromagnetic bandgap-backed antenna suppresses surface waves and minimizes back radiation, improving efficiency and gain for wearable device applications \cite{iqbal2019electromagnetic}.}
    \label{fig:metasurfaces}
\end{figure*}

DMAs can benefit tri-hybrid MIMO through high-precision beam steering with low power consumption. The performance depicted in Fig. \ref{fig:DMA based tri-hybird simulation results} and the various configurations shown in Fig. \ref{fig:metasurfaces} suggest that DMAs can substantially contribute to this hybrid framework by offering an energy-efficient solution for frequency-agile beamforming. Additionally, the functional characteristics outlined below further demonstrate that metasurface antennas are strong candidates for tri-hybrid systems.

One of the key strengths of metasurface antennas is their ability to achieve polarization reconfiguration, which further enhances their versatility in hybrid systems. By using the anisotropy of the unit cell designs, metasurfaces can support multiple polarization states. For example, the design in \cite{zhu2014design} shows that linear, left-handed, and right-handed circular polarizations can be realized by rotating the metasurface structure. Furthermore, dual-polarization is made possible with designs like the one in \cite{wang2022dual}, where orthogonal varactor diode pairs allow independent beam steering for vertical and horizontal polarizations. This dual-polarization control is highly advantageous for massive MIMO systems, where both polarization flexibility and compactness are critical requirements.

Furthermore, metasurface antennas excel in beam steering, making them ideal for applications requiring agile and dynamic beam control. A design introduced in \cite{xu2021compact} uses a metasurface lens array antenna that offers low-cost, high-gain beam steering. By dividing a large lens into smaller elements, the design minimizes antenna thickness without sacrificing performance. Beam steering is achieved \edit{by dynamically switching feed positions across different regions of the metasurface lens array} with phased arrays providing fine control over the beam direction. These characteristics make metasurfaces an excellent fit for tri-hybrid architectures requiring precise, responsive beam control.

Metasurface antennas are capable of frequency reconfiguration, further supporting their suitability in tri-hybrid designs where multi-band operation is essential. The dual-layer metasurface in \cite{liu2024dual} achieves frequency tuning across 24 GHz radar and 5G bands, with a notable shift of 750 MHz for every 1~mm adjustment in feeder-metasurface distance. This compact design replaces the need for multiple antennas dedicated to different frequency bands, such as those proposed in \cite{KangWireless2024}, providing a more efficient and adaptable solution for sub-6~GHz, mmWave, and even sub-THz bands.

\subsection{Parasitic element-assisted antenna}

Tunable passive elements arranged around an active antenna harness parasitic coupling to achieve dynamic reconfigurability. For example, dynamic scattering arrays -- as demonstrated in \cite{dardari2024DSA} -- can generate high-directivity multibeams within a compact footprint,  smaller than that required by conventional antenna arrays. This approach is part of a broader strategy known as \emph{electromagnetic precoding}, which offers a versatile framework for shaping electromagnetic fields across diverse hardware architectures. Unlike a simple reconfigurable antenna, a collection of active elements and tunable passive elements are one way to realize the multi-port vision in Fig.\ref{fig:mimo_architectures}(c). The multi-port nature comes because the parasitic elements impact the collective patterns generated from the multiple active sources. In this section, the term parasitic element-assisted antenna extends beyond the conventional definition of parasitic elements. While traditional parasitic antennas rely on passive elements directly coupled to active radiators, we consider a broader category that encompasses various auxiliary units that assist antenna operation, including reconfigurable metastructures, dynamic scattering elements, and metasurface-based reflectors and transmitarrays. These structures, though not physically integrated with the radiating element in a classical sense, effectively function as parasitic layers that contribute to beamforming, polarization control, and frequency manipulation. As such, this perspective aligns with the multi-layer electromagnetic processing paradigm, where passive and active elements are jointly optimized to enhance the overall system performance. Although the focus in our paper is on reconfigurable antennas as a specific application of electromagnetic precoding, the underlying principles naturally extend to more complex configurations, such as dynamic scattering arrays and multi-port reconfigurable surfaces.

\begin{figure*}[!t]
    \centering
    \begin{subfigure}[b]{0.3\textwidth}
        \centering
        \includegraphics[width=\textwidth]{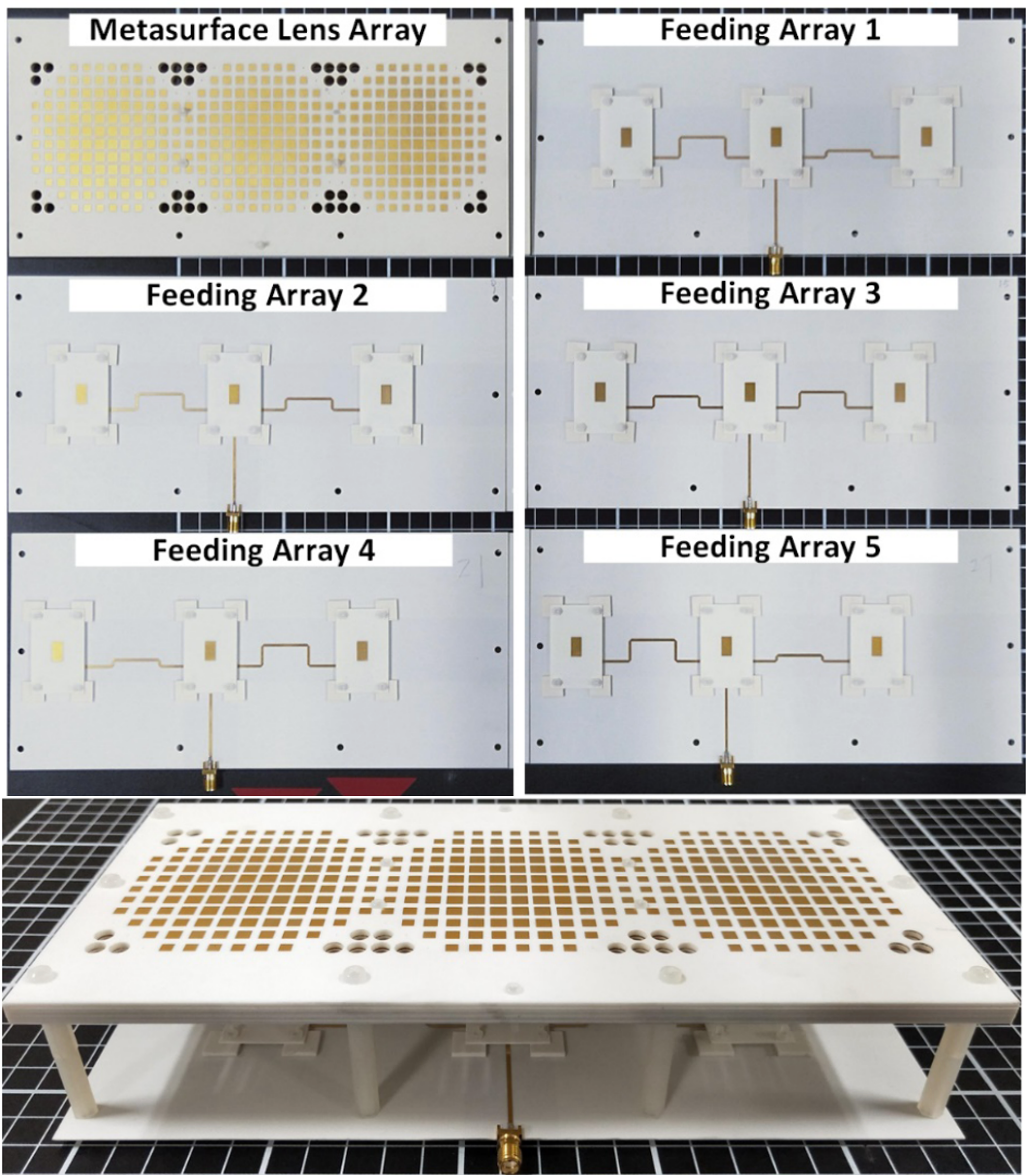}
        \caption{Transmitarray using metasurface \cite{xu2021compact}}
        \label{fig:subfig5}
    \end{subfigure}
    \hspace{0.5cm}
    \begin{subfigure}[b]{0.3\textwidth}
        \centering
        \includegraphics[width=\textwidth]{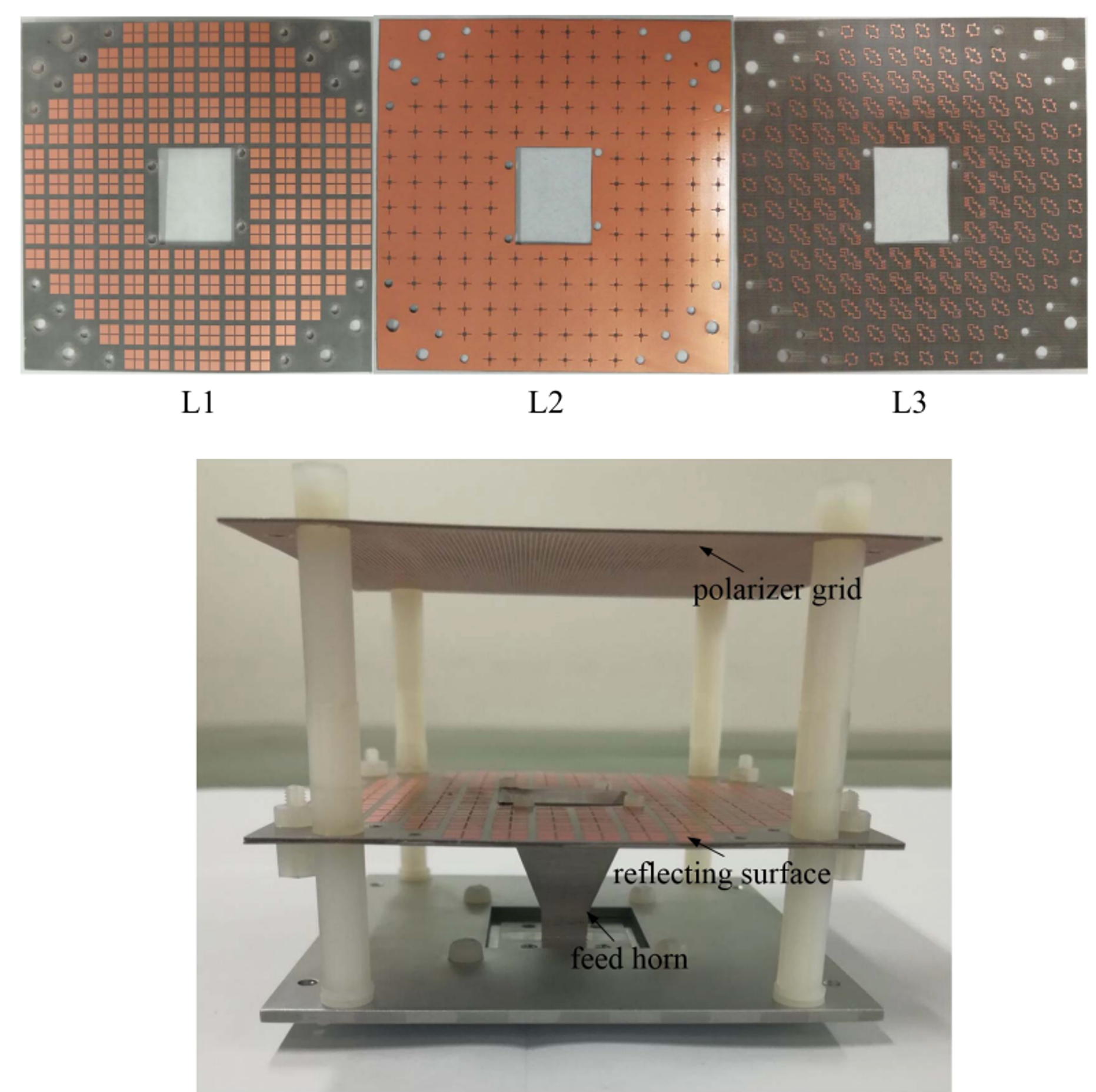}
        \caption{Reflectarray using metasurface \cite{cao2020novel}}
        \label{fig:subfig6}
    \end{subfigure}
    \hspace{0.5cm}
    \begin{subfigure}[b]{0.3\textwidth}
        \centering
        \includegraphics[width=\textwidth]{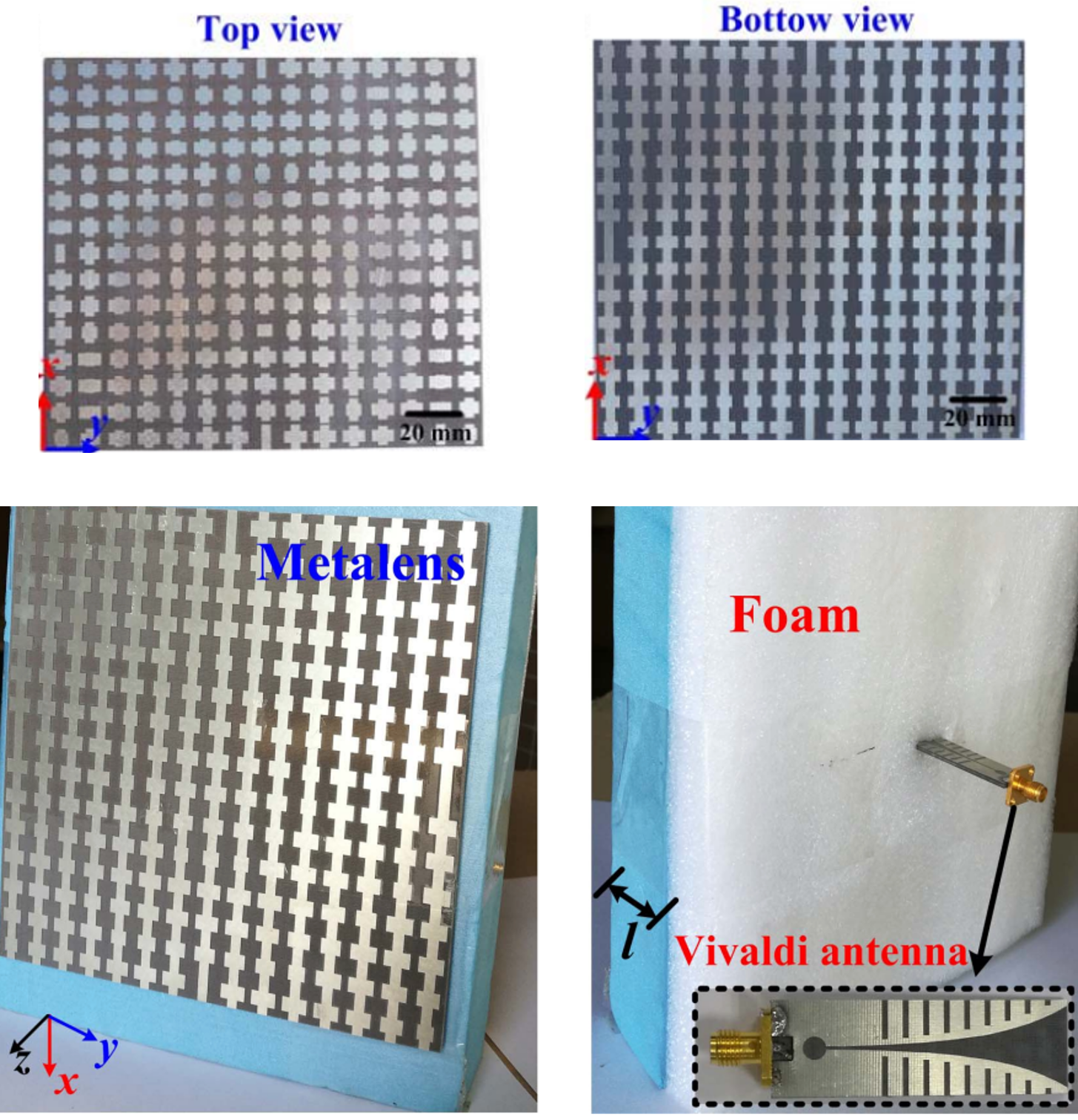}
        \caption{Multifunctional metasurface enabling both reflectarray and transmitarray operations \cite{cai2018high}}
        \label{fig:subfig7}
    \end{subfigure}
    
    \caption{Various embedded reconfigurable structure designs.
    (a) A transmitarray metasurface focuses and directs transmitted waves to enhance gain and beam control \cite{xu2021compact}. 
    (b) A reflectarray metasurface manipulates reflected waves to achieve precise beam steering \cite{cao2020novel}. 
    (c) A multifunctional metasurface integrates transmitarray and reflectarray functionalities, enabling versatile wave manipulation \cite{cai2018high}.}
    \label{fig:metasurface_parasitic_layers}
\end{figure*}

Metasurfaces can serve not only as antennas themselves but also as auxiliary devices in existing antenna communication systems. For instance, metasurface-based reflectors improve coverage by reflecting incident waves in specific directions, while transmitarrays enable highly directive beams. These external parasitic element-assisted antennas are often easier to implement than fully integrated metasurface antennas and do not require significant changes to the transceiver structure, making them highly compatible with existing systems~\cite{hum2014reconfigurable}. Fig.~\ref{fig:metasurface_parasitic_layers} illustrates examples of such implementations, highlighting how metasurfaces can be used indirectly to enhance the performance of traditional antenna systems.

When it comes to beam shaping and enhancing antenna performance, metasurface transmitarrays are particularly effective. The design in \cite{xu2021compact} uses a tunable lens composed of smaller elements, significantly reducing antenna thickness while maintaining the desired focal ratio. This approach achieves beam steering through feed antenna movement combined with a low-cost phased array, providing $\pm 30^\circ$ beam coverage and delivering a high 19.1 dBi gain with low sidelobe levels. Other implementations, such as the silicon metasurface used in \cite{headland2016dielectric}, demonstrate beam focusing capabilities at 1 THz, while the anisotropic metasurface in \cite{yang2018design} enables simultaneous beamforming and beam combining at different angles. Such features make these designs ideal for MIMO systems, which require fewer feed antennas while still achieving efficient beam control.

Stacked intelligent metasurfaces (SIMs) are an advanced application of metasurfaces in wireless sensing and communication. SIMs perform signal processing directly in the EM domain, leveraging the layered structure of metasurfaces. 
Each meta-atom within the layers interacts with incoming electromagnetic waves, modifying their properties and re-radiating them. This re-radiation from multiple meta-atoms occurs simultaneously, enabling parallel processing of the waves as they propagate through the layers.
This wave-domain processing eliminates the need for complex digital signal processors, significantly reducing hardware complexity and achieving ultra-fast computation speeds. For example, as demonstrated in \cite{SIMcomputing_1720551601}, SIMs are capable of replacing traditional digital precoding in MIMO systems, where wave-domain beamforming can effectively suppress inter-stream interference while lowering the reliance on high-resolution DACs. Additionally, their real-time programmability enables dynamic tuning to environmental changes, optimizing functions such as beamforming and channel estimation. This results in reduced latency and energy consumption, making SIMs a possible solution for next-generation communication systems requiring real-time adaptability and efficiency.

Another important advantage of parasitic element-assisted antennas is their ability to control polarization. Metasurfaces can act as reconfigurable polarizers when placed near transceiver arrays, allowing for flexible control of polarization states. The dual-mode transmissive metasurface described in~\cite{xu2017dual} enables independent adjustment of orthogonal polarizations, providing greater flexibility in polarization reconfiguration. Similarly, \cite{yang2021folded} presents a folded transmit and receive array based on a metasurface for circular polarization, which reduces the overall size of the antenna without sacrificing performance. Other designs, such as the broadband folded antenna in \cite{li2021broadband}, use an ultrathin metasurface to achieve efficient transmission, phase compensation, and 90° polarization rotation. These metasurface-assisted polarizers enable tunable polarization control, which is especially beneficial for communication systems that do not rely on a particular MIMO array configuration. High gains and low cross-polarization levels demonstrated in \cite{cai2018high} further highlight the practicality of metasurfaces for polarization reconfiguration.

Metasurfaces are also highly effective for frequency reconfiguration in communication systems. Time-modulated metasurfaces, as demonstrated in \cite{taravati2021pure}, allow spurious-free, linear frequency conversion through dynamically controlled supercells composed of both static and time-varying resonators. The surface-interconnector-phaser-surface (SIPS) architecture acts as a multiband filter to suppress unwanted harmonics while providing real-time frequency control via external signals, such as FPGA. This architecture enables large frequency shifts, which are critical in telecommunications. For example, \cite{taravati2021pure} shows successful up-conversion from 2.33~GHz to 3.39~GHz and down-conversion from 5 GHz to 3.21 GHz with strong spurious signal suppression, demonstrating that metasurfaces offer a low-profile, efficient solution for systems requiring dynamic frequency control.

\subsection{Structurally reconfigurable arrays} 
Structurally reconfigurable arrays modify the physical geometry of the device to change antenna properties. As a relatively new area of research, these arrays are still in the early stages of development but are gaining significant attention for their potential in dynamic adaptability. Unlike metasurface antennas, which primarily manipulate electromagnetic properties, structurally reconfigurable arrays achieve reconfiguration through direct alterations in the physical structure of the antenna, including methods such as fluid-based reconfiguration. 

Recently, flexible and fluid (movable) antennas have garnered significant attention for their ability to achieve dynamic reconfiguration by altering the structure of the antenna itself \cite{New_fluid, Flexible_1720549736, huang2021liquid, shen2020beam}. These antennas are generally categorized into three types:  
\paragraph{Fluid-driven element arrays} 
These antennas utilize the movement of a conductive or dielectric fluid within a controlled medium to dynamically adjust the effective placement or orientation of radiating elements. Rather than physically displacing solid antenna elements, the fluid modifies electromagnetic properties such as capacitance or resonance, leading to changes in the radiation pattern.

\paragraph{Liquid-metal-based radiators}
This category includes antennas that employ conductive fluids, such as liquid metal or ionized solutions, to form reconfigurable radiating elements. The ability to dynamically reshape these liquid conductors enables tunability in both frequency and polarization, making them suitable for applications requiring agile electromagnetic responses.

\paragraph{Deformable structural arrays}
These antennas leverage flexible and stretchable materials, such as shape-memory alloys or elastomer substrates, to physically alter the array structure. By bending, folding, or stretching the antenna geometry, these designs enable dynamic control over radiation characteristics, allowing adaptation to varying communication scenarios.

These antenna designs remain an active area of research, with each configuration offering its own unique advantages and inherent trade-offs.

For example, the first type of antenna proposed in~\cite{shen2020beam} employs a surface wave fluid antenna that dynamically adjusts the position of a fluid metal radiator, such as Galinstan, to control the radiation direction. This enables precise beam steering while eliminating the need for multiple radiating elements or complex feeding networks, thereby simplifying the design. By leveraging the non-radiating nature of surface waves, the antenna reduces radiation loss and enables effective beam steering with a single RF input. This design not only simplifies the feeding mechanism but also maintains spatial diversity, as the fluid metal radiator can adjust the main beam direction dynamically without requiring multiple antennas. However, the reliance on liquid metal introduces challenges in long-term stability and response time, particularly in high-speed reconfiguration scenarios. Moreover, fabrication complexity and potential reliability issues associated with liquid metal movement remain key obstacles to widespread adoption.

The second type, a frequency- and polarization-reconfigurable slot antenna utilizing liquid metal, was introduced in \cite{liu2020frequency}. This design employs microfluidic channels embedded within a polydimethylsiloxane (PDMS) structure, allowing liquid metal injection to dynamically change the antenna properties. By controlling the liquid metal distribution, the antenna achieves three distinct polarization states: linear polarization (LP), left-hand circular polarization (LHCP), and right-hand circular polarization (RHCP). Additionally, the antenna offers tunable frequency bands for each polarization state, with a 3 dB axial ratio bandwidth (ARBW) of up to 26.42\%. The use of liquid metal as a reactive loading avoids the limitations associated with traditional RF switches, such as nonlinearity and low power handling, while maintaining efficiency and compactness. However, practical deployment remains limited due to fabrication difficulties and the need for precise microfluidic control. The potential for leakage or oxidation of the liquid metal also raises reliability concerns in long-term applications, making widespread implementation challenging.

The third type, flexible antenna arrays, as described in~\cite{Flexible_1720549736}, incorporates a rotatable, bendable, and foldable skeleton that allows for dynamic reconfiguration. These arrays are designed to adapt their physical shape, enabling a range of configurations to meet varying performance demands. By altering the position and orientation of the antenna elements, the system achieves flexibility in optimizing its operation for different scenarios. Experimental results demonstrate that such structural changes have a direct impact on the sum-rate performance, particularly during transitions between configurations. While this adaptability makes the design promising for environments requiring frequent adjustments, the mechanical durability of the flexible structures and their long-term reliability under repeated deformation remain critical challenges. Additionally, integrating flexible materials with conventional RF components requires further refinement to minimize signal degradation and ensure consistent performance.

\section{Tri-hybrid MIMO architecture}
\label{sec_tri_hybrid}


In this section, we introduce and explain more fully how reconfigurable antennas can be incorporated into existing MIMO architectures to form a tri-hybrid system. The term tri-hybrid refers to the fact that signal processing is separated into three distinct domains: digital, analog, and antenna. We begin by reviewing the hybrid analog-digital MIMO solution. Then we cover possible arrangements for tri-hybrid MIMO. Finally, we describe the benefits of different types of reconfigurable antennas.

\subsection{MIMO with hybrid analog-digital precoding}
Hybrid precoding architectures are MIMO systems where in general the number of digital ports is less than the number of antennas. Some form of analog beamforming -- using phase shifters, switches or other components -- is performed to map the digital signals to /from the larger number of antennas. Without loss of generality, we describe a transmit hybrid architecture in the context of a MIMO-OFDM (orthogonal frequency division multiplexing) system with a single user and then show how the model changes with the introduction of reconfigurable antennas. We focus on tri-hybrid capabilities at the transmitter, but the model can be naturally used for the receiver as well. We provide expressions for the MIMO channel and the received signal, under some assumptions, as part of the performance analysis in Section \ref{sec_performance}. 

We describe the complex baseband equivalent signal model for the hybrid architecture as in~\cite{HeathEtAlOverviewSignalProcessingTechniques2016}. The transmitter, which is equipped with $\Nt$ antennas and $\Ntrf$ RF chains, communicates with an $\Nr$-antenna receiver. \edit{The number of RF chains corresponds to the number of antenna ports.} The transmitter sends $\Ns$ data streams over subcarriers $k=1,...,K$. We let $\bsfs[k]$ denote the $\Ns$-length symbol vector. We assume that $\bsfs[k]$ is zero-mean and unit-variance. In a hybrid architecture, $\bsfs[k]$ is first processed by an $\Ntrf \times \Ns$ frequency-selective digital precoder $\Fdig[k]$ and then an $\Nt \times \Ntrf$ frequency-flat analog precoder $\Fana$. We assume a phase-shifter-based analog precoder, which imposes the constant-modulus constraint on the precoder entries as $\left( \Fana \Fanac \right)_{n,n} = 1/ \Nt$.
The \edit{discrete-time complex baseband} input to the antenna array is the transmit signal $\bsfx[k] \in \C^{\Nt}$
\begin{equation}
    \label{eq_tx_signal}
    \bsfx[k] = \Fana \Fdig[k] \bsfs[k], \quad k=1,\ldots,K.
\end{equation}
The $\Nt \times \Ns$ hybrid precoder $\Fhyb[k] = \Fana \Fdig[k]$ in \eqref{eq_tx_signal} divides processing between the analog and digital domains. This split reduces the number of RF chains in MIMO devices with many antennas. Fully-digital architectures, while effective, consume inordinate amounts of power because of each dedicated RF chain. A hybrid architecture satisfies $\Ntrf < \Nt$, with the architecture referred to as fully-analog in the extreme case $\Ntrf = 1$. 

The connection between the digital and analog processing blocks in hybrid systems can be tailored to reduce hardware complexity and power consumption. In a fully-connected architecture, the analog precoder feeds all the antennas and the structure of $\Fana$ is only limited by the \edit{possible} modulus constraint imposed by the phase shifter architecture. Another option is to use each RF chain to feed a subset of antennas. We assume that $\Nt$ is divisible by $\Ntrf$ such that the array can be divided into $\Nsub = \Nt / \Ntrf$ subarrays of equal size. For the $\ell$th subarray, let $\bsff_{\ana, \ell}$ denote the phase shifter beamformer. In the partially-connected architecture, the analog precoder exhibits a block diagonal form as
\begin{equation}
\label{eq_partial_ana_precoder}
      \Fana = \blkdiag\left( \bsff_{\ana, 1}, \, \cdots, \, \bsff_{\ana, \Nt}\right),
\end{equation}
where the beamforming vectors $\bsff_{\ana, \ell}$ are also subject to the modulus constraint.
Partial connectivity in the hybrid architecture simplifies hardware implementations by reducing the number of RF paths. The partially-connected architecture also serves as a good point of comparison for tri-hybrid MIMO architectures that make use of reconfigurable antennas. 

\subsection{Incorporating reconfigurable antennas to realize the tri-hybrid architecture}

One of the main benefits of incorporating a new layer of reconfigurable antennas is to consume less power. In general, reconfigurable antenna designs use low-power components, such as PIN diodes and tunable meta-elements, to switch between configurations. These reconfigurable components consume much less power than traditional RF processing tools such as phase shifters. For example, DMAs, which are a type of pattern reconfigurable antenna, achieve per-element beamforming through PIN or varactor diodes that consume negligible power compared to phase shifters \cite{CarlsonEtAlHierarchicalCodebookDesignWith2023}. The power consumption benefits of a hybrid architecture can be improved by shifting the signal processing burden of the system from analog processing to the reconfigurable antennas. For example, a partially connected hybrid architecture with $16$ antennas per branch could be replaced by an interconnection of two DMAs each that does the job of $8$ antennas. In this way, the number of phase shifters and associated components would be reduced. 

A second benefit of  incorporating a new layer of reconfigurable antennas is to expand the supported aperture size. For example, consider a reconfigurable antenna like a DMA that can take the replace of a subarray of $8$ antennas. Connecting such a reconfigurable antenna to a hybrid system with $\Nt=100$ antennas scales the effective aperture (in terms of the number of radiators) to $8 \times 100 = 800$. This application of the tri-hybrid architecture may be very important in scaling up the arrays from 5G to support the applications of 6G, e.g. to arrays with thousands of antennas that  may be used in the upper midband \cite{KangWireless2024}

A third benefit of incorporating a new layer of reconfigurable antennas as found in the tri-hybrid MIMO architecture is the flexibility that become available from reconfigurability. Polarization reconfigurable antennas, for example, can combat channel depolarization and achieve throughput gains in both line-of-sight and non-line-of-sight settings \cite{CastellanosHeathLinearPolarizationOptimizationWideband2023} without the need for antennas with two polarization. Frequency reconfigurable antennas also enable new avenues to support higher bandwidths in wideband MIMO \cite{ObeidatEtAlDesignFrequencyReconfigurableAntennas2010}. Pattern reconfigurability gives another dimension of flexibility to accomplish objectives like reducing transmissions to or receptions from spectrum incumbents \cite{HasBahCet_DownlinkMUMIMORA_18}. 
Overall, reconfigurability offers the system a novel way to adapt to dynamic channel conditions.


\begin{figure*}[!t]
    \centering
\includegraphics[width = 0.8\textwidth]{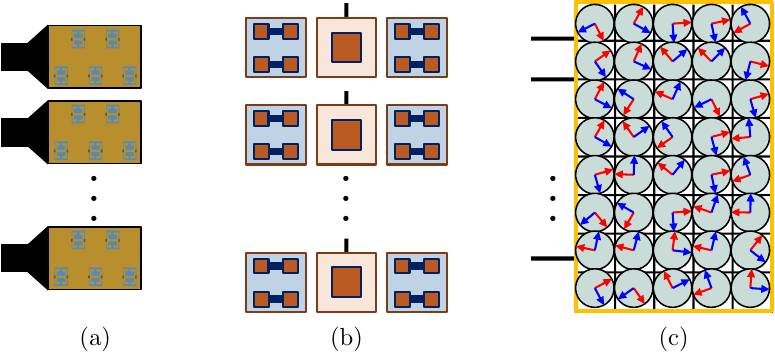}
\caption{Tri-hybrid antenna arrays with (a) dynamic metasurface antennas, (b) parasitic antenna arrays, and (c) polarization reconfigurable antennas. Each array exhibits different precoding structures.}
\label{fig_tri_hybrid_configurations}
\end{figure*}

\subsection{Re-thinking precoding in the context of the tri-hybrid MIMO architecture}

The reconfigurable antenna design impacts both the precoding and array structure. Fig.~\ref{fig_tri_hybrid_configurations}, we show three different kinds of reconfigurable arrays: a DMA, a tunable parasitic antenna array, and a polarization reconfigurable array. Each design achieves reconfiguration in a different way. In the previous section, we highlighted that reconfigurable antennas are a way to reduce the number of analog and digital chains to save power, or a means to expand the aperture without expanding the digital / analog connectivity or as a tool to achieve a higher layer of flexibility. It is important to formulate a mathematical expression that can model many kinds of reconfigurability and makes it possible to achieve the different benefits as part of the larger MIMO architecture.

To start this development, we need to build another layer of notation to help differentiate between the number of antenna ports and the number of radiating elements. The term $
\Nt$ refers to the number of antennas in the conventional hybrid architecture. In the tri-hybrid system, the additional layer of reconfigurability contributes to the true number of radiating elements $\Ntrad$ being larger than $\Nt$. DMA waveguides, for example, are fed by RF signal and tunable slots are used to configure the gain pattern. Assuming each DMA has $\Ntuc$ slots, there are $\Ntrad = \Nt \Ntuc$ radiating elements. Similarly, parasitic arrays also expand the effective array by adding tunable passive antenna elements that couple to active antennas. Assuming that each active antenna corresponds to $\Ntpar$ parasitic antennas, the number of radiating elements is $\Ntrad = \Nt (\Ntpar + 1)$ 
In the case of the polarization reconfigurable case, the array can be modeled as the superposition of two orthogonally polarized arrays with $\Ntrad = 2 \Nt$ \cite{CastellanosHeathLinearPolarizationOptimizationWideband2023}. In general, the effect of the reconfigurable antenna can be modeled through a $\Ntrad \times \Nt$ precoding matrix $\Fant[k]$. This yields a length $\Ntrad$ antenna transmit signal in terms of a $\Ntrad \times \Ns$ tri-hybrid precoder $\Ftri[k]$ as
\begin{equation}
    \label{eq_ant_signal}
    \bsfx_{\ant}[k] = \Fant[k] \Fana \Fdig[k] \bsfs[k], \quad k=1,\ldots,K.
\end{equation}
The effective array dimensions have important ramifications for both channel modeling and signal design.

In the tri-hybrid architecture, the effective precoder 
is the product of three components $\Fant[k] \Fana \Fdig[k]$, each a tall matrix (contributing to dimensionality reduction). The antenna precoder, or electromagnetic precoder, $\Fant[k]$ is of dimension $\Ntrad \times \Nt$.  is frequency-selective with index $k$ because of the innate frequency response of the array elements. In some reconfigurable array designs, like parasitic arrays, it may be common though to treat it as frequency flat. The constraints on $\Fant[k]$ depend on the specific reconfigurable antenna or surface. The analog precoder $\Fana$ is of dimension $\Nt \times \Ntrf$. It connects the digital ports to analog ports. While it can also be modeled as frequency selective, e.g. when using true-time delay precoding \cite{LongbrakeTrueTimeDelayBeamsteering2012}, 
it is normally assumed to be frequency flat as it captures the operation of (ideal) phase shifters. The analog precoder often has unimodular or finite phase constraints that account for the use of discretized phase shifters. Finally, $\Fdig[k]$ is the $\Ntrf \times \Ns$ digital precoder that maps $\Ns$ data streams to $\Ntrf$ ports. As $\Fdig[k]$ is implemented as part of the digital signal processing, it many not have hardware-dependent constraints aside from those implied by the power constraint. 

The transmit power constraint -- especially where it is placed -- needs to be revisited in the tri-hybrid precoding architecture. Wireless systems normally constrain the maximum radiated power in the allocated frequency band for user safety and interference through measures like the effective isotropic radiated power (EIRP) \cite{IECDeterminationRFFieldStrength2022}. 
Treating the antennas as isotropic radiators, then this becomes a constraint placed on $\bsfx[k]$ \cite{HeathLozanoFoundationsMIMOCommunications2018}. In the hybrid architecture, with power limit $P$, this gives the constraint $\sum_{k=1}^K \norm{\Fana \Fdig[k]}_{\fro}^2 \leq P$. We refer to this constraint as an \edit{antenna input power (AIP)} constraint because it acts on the antenna input port signal. The AIP constraint is simple but fails to account for the reconfigurable antenna characteristics that further translate the signals from the antenna ports to the radiating elements. For example, DMAs may not radiate all of the power injected into the waveguide and the radiated power can change depending on the DMA slot configurations. By suitably modeling the manifold of values $\Fant[k]$, a constraint can be placed on the \edit{antenna output power (AOP)} as $\sum_{k=1}^K \norm{\Fant[k] \Fana \Fdig[k]}_{\fro}^2 \leq P$. The AOP can further be used to define a radiated power constraint by incorporating the antenna losses and element patterns. The antenna properties are directly embedded into the constraint through $\Fant[k]$ to account for the reconfigurable array radiation efficiency. In summary, when doing tri-hybrid precoding, the power constraint may be placed at the input or output. 

The specific design for the reconfigurable antenna determines the structure of $\Fant[k]$. For example, a DMA may be modeled as a linear array but with beamforming weights that are constrained by a Lorentzian constraint, which couples the phase and magnitude response \cite{CarlsonEtAlHierarchicalCodebookDesignWith2023, SmithEtAlAnalysisWaveguideFedMetasurface2017}. DMAs also result in result in unequal output power in each radiating slot because the slots at the end of each antenna see weaker excitations due to the waveguide attenuation. Similarly, parasitic arrays are limited by the passive reconfiguration of each reconfigurable element, where the overall array pattern can only be adjusted by changing the parasitic antenna impedances. 
Signal processing for the polarization reconfigurable array takes a different form from the other two examples. \edit{Polarization precoding can be seen as an operation over an effective unpolarized channel with $\Ntrad = 2\Nt$ as shown in \cite{CastellanosHeathLinearPolarizationOptimizationWideband2023}}. These examples showcase that tri-hybrid MIMO models are not straightforward extensions of hybrid MIMO, as each reconfigurable antenna implementation results in a unique architecture. A full explanation of all possibilities is a topic for future work. In the following section, we analyze the tri-hybrid architecture with DMAs in detail and compare its rate performance and power consumption to other existing architectures.

\section{Performance comparison between MIMO architectures}
\label{sec_performance}


\begin{figure*}
    \centering
\begin{tabular}{cc}
\includegraphics[width = 0.45\textwidth]{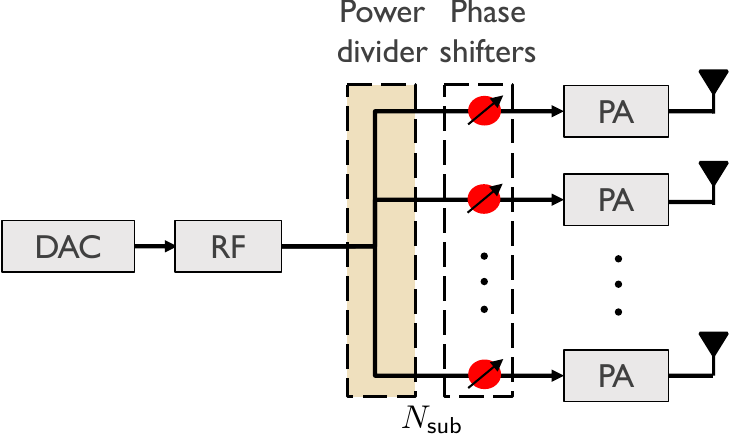} & \includegraphics[width = 0.45\textwidth]{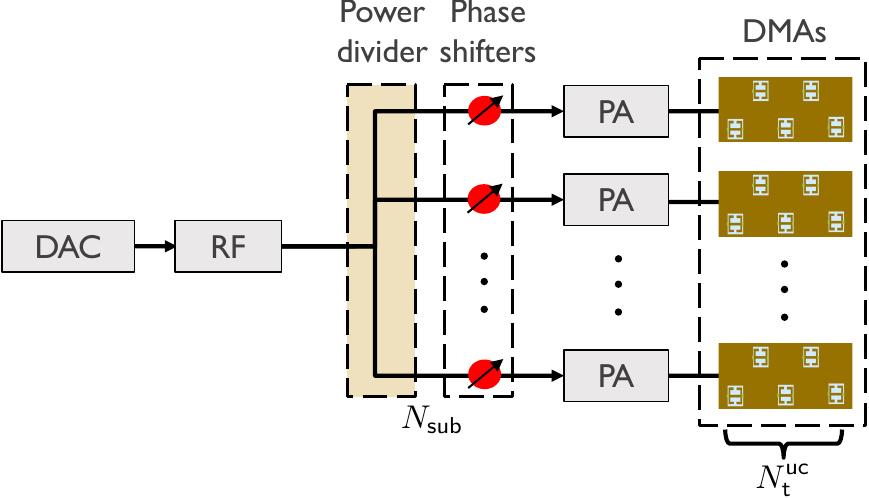} \\
(a) Hybrid MIMO subarray & (b) Tri-hybrid MIMO subarray
\end{tabular}
\caption{Subarray diagram for both the hybrid and tri-hybrid architectures. (a) In the hybrid architecture, each RF chain connects to $\Nsub = \Nt / \Ntrf$ phase shifters to feed the antennas. (b) The tri-hybrid architecture replaces the antenna array with DMAs for additional EM precoding, each of which contains $\Ntuc$ radiating slots.}
\label{fig_hybrid_subarrays}
\end{figure*}

Tri-hybrid MIMO expands on the traditional hybrid MIMO architecture to  different performance benefits compared to traditional hybrid architectures. In this section, we consider a tri-hybrid architecture that leverages DMAs as shown in Fig. \ref{fig_hybrid_subarrays}. 
We select DMAs because their behavior has been well characterized in prior works \cite{CarlsonEtAlHierarchicalCodebookDesignWith2023, SmithEtAlAnalysisWaveguideFedMetasurface2017}, making them a suitable example for analyzing the constraints imposed on $\Fant[k]$. Specifically, DMAs enforce Lorentzian-type constraints that couple phase and amplitude response, shaping the achievable beamforming patterns. Additionally, their waveguide-fed structure enables low-power reconfigurable beamforming, as they do not require a dedicated phase shifter for each element, unlike conventional hybrid architectures. This unique structure results in a favorable trade-off between power consumption and beamforming flexibility. We contrast this design against traditional hybrid architectures, noting that the varying numbers of RF chains and phase shifters in each system create tradeoffs between spectral efficiency and power consumption. Finally, we model component loss and power consumption for each precoding architecture and conclude the section with numerical comparisons of the different MIMO solutions to demonstrate the benefits of tri-hybrid MIMO in terms of energy efficiency. 

\subsection{Tri-hybrid precoding}

\edit{The tri-hybrid architecture in this section replaces static antennas with reconfigurable DMAs.} We assume a DMA subarray architecture in which each antenna feed excites $M$ DMAs. We let the number of unit cells per DMA be $\Ntuc$ such that the total number of radiators in the array is $\Ntrad = \Nt M \Ntuc$ antenna elements. We assume the DMAs are frequency-flat over the system bandwidth and denote the DMA-based precoder as the $\Ntrad \times \Nt$ frequency-flat matrix $\Fdma$. The DMA transmit signal after tri-hybrid precoding is  
\begin{equation}
  \label{eq:tx-dma-signal}
  \bsfx_{\dma}[k] = \Fdma \bsfx[k] = \Fdma \Fana \Fdig[k] \bsfs[k], \quad k=1,\ldots,K.
\end{equation}
The DMAs naturally create subarrays of radiators because each waveguide only has a single input port. This places a block-diagonal constraint on $\Fdma$. Let the length $M \Ntuc$ vector $\bsff_{\dma,n}$ be the $n$th DMA subarray beamformer. The DMA precoder is then
\begin{equation}
  \label{eq_dma_precoder}
  \Fdma = \blkdiag\left( \bsff_{\dma, 1}, \, \cdots, \, \bsff_{\dma, \Nt}\right).
\end{equation}
The same partially-connected structure is shown in both \eqref{eq_partial_ana_precoder} and \eqref{eq_dma_precoder}. The key difference between the two precoders is the additional constraints placed on $\Fdma$ arising from the nature of the passive reconfigurable slots.

The DMA precoder constraints are inherited from characteristics of the leaky-wave slot responses \cite{SmithEtAlAnalysisWaveguideFedMetasurface2017}. Each slot contains a reconfigurable component, such as a varactor or PIN diode, that tunes the slot resonance frequency. The impact of this on the slot response can be modeled through a beamforming weight $\alpha$ that satisfies a Lorentzian constraint a~\cite{ShlezingerEtAlDynamicMetasurfaceAntennasUplink2019}
\begin{equation}
  \label{eq:Lorentzian-constraint}
  \cL = \left\{\alpha = \frac{-j + e^{j\varphi}}{2}: \varphi \in [0,2\pi) \right\}.
\end{equation}
We let $\alpha_{m,n} \in \cL$ denote the DMA weight applied by the $m$th unit cell of the $n$th waveguide. The constraint couples the phase shift and gain applied by the weight and is a key feature of the DMA. In addition to the Lorentzian response, the slots see slightly different inputs due to phase advance along the waveguide. We characterize this effect through a waveguide channel term $\eta_{m,n}$ with $\abs{\eta_{ m,n}} \leq 1$. Combining the waveguide channel and the Lorentzian weight effect, the $n$th DMA beamformer is 
\begin{equation}
  \label{eq_dma_beamformer}
  \bsff_{\dma, n} = \left[ \eta_{1, n} \alpha_{ 1, n}, \, \cdots, \, \eta_{\Ntuc, n} \alpha_{\Ntuc, n} \right]^{\T}.
\end{equation}
We denote the set of feasible DMA precoders that satisfy \eqref{eq_dma_precoder} and \eqref{eq_dma_beamformer} as $\cF_{\dma}$.

\edit{We assume a fully-digital $\Nr$ antenna receiver, perfect channel knowledge and perfect frequency offset and carrier synchronization to simplify the analysis.} We let the $\Nr \times \Ntrad$ matrix $\bsfH[k]$ be the wireless channel between the DMA slots and the receive antenna. We also define $\bsfn[k] \in \C^{\Nr}$ as the additive noise. The received signal at the $k$th subcarrier is
\begin{equation}
  \label{eq:io-model}
  \bsfy[k] = \bsfH[k] \Fdma \Fana \Fdig[k] \bsfs[k] + \bsfn[k].
\end{equation}
We assume a per-entry noise variance $\sigma^2$ and that $\bsfn[k]$ is circularly-symmetric Gaussian noise with distributed as $\cC \cN(\bm{0}, \bI_{\Nr} \sigma^2)$.

The transmitter designs the tri-hybrid precoders to maximize spectral efficiency. In this case, it is sufficient to maximize the mutual information with post-DMA power constrained by $P$ as
\begin{IEEEeqnarray}{rl}
  \label{eq:tri-hyb-opt}
 \max_{\Fdig, \Fana, \Fdma} \quad & \sum_{k=1}^K \log_2 \abs{\bI_{\Nr} + \frac{1}{\sigma^2} \bsfH[k] \Fdma \Fhyb[k] \Fhybc[k] \Fdmac \bsfH^*[k]} \nonumber \\
 \text{s.t} \quad & \sum_{k=1}^K \norm{\Fdma \Fhyb[k]}_{\fro}^2 \leq P,  \\
 & \Fana \in \cF_{\ana}, \nonumber \\
 & \Fdma \in \cF_{\dma}. \nonumber
\end{IEEEeqnarray}
Due to the complexity of jointly optimizing all precoders, we simplify the problem by optimizing each precoding matrix separately.

We first fix the DMA precoder and use the results in \cite{ParkEtAlDynamicSubarraysHybridPrecoding2017} to optimize the hybrid precoder. We assume that the hybrid precoder $\Fhyb[k] = \Fana \Fdig[k]$ is unconstrained and optimal such that waterfilling through the hybrid precoder is possible. Let the effective hybrid channel be $\Heffhyb[k] = \bsfH[k] \Fdma (\Fdmac \Fdma)^{-1/2}$ and $\Feffhyb[k] = (\Fdmac \Fdma)^{1/2} \Fhyb[k]$. The optimal effective hybrid precoder for a fixed $\Fdma$ is then
\begin{IEEEeqnarray}{rl}
 \max_{\Feffhyb} \quad & \sum_{k=1}^K \log_2 \abs{\bI_{\Nr} + \frac{1}{\sigma^2} \Heffhyb[k] \Feffhyb[k] \Feffhyb^*[k] \Heffhyb^*[k]}\nonumber \\
 \text{s.t} \quad & \sum_{k=1}^K \Vert \Feffhyb \Vert_{\fro}^2 \leq P, 
   \label{eq:tri-hyb-opt-alt}
\end{IEEEeqnarray}
which is obtained by waterfilling. The effective hybrid channel is partitioned as
\begin{equation}
  \label{eq:1}
  \Heffhyb[k] = \left[ \frac{\bsfH_1[k] \bsff_{\dma, 1}}{\abs{\bsff_{\dma,1}}}, \, \cdots, \, \frac{\bsfH_{\Nt}[k] \bsff_{\dma, \Nt}}{\abs{\bsff_{\dma, \Nt}}} \right]
\end{equation}
to define the $n$th subarray channel covariance matrix
\begin{equation}
  \label{eq:sub-channel-cov}
  \bsfR_n = \frac{1}{K} \sum_{k=1}^K \bsfH_n^*[k] \bsfH_n[k].
\end{equation}
In \cite{ParkEtAlDynamicSubarraysHybridPrecoding2017}, the partially-connected analog precoder is computed through the maximal eigenvectors of $\bsfR_n$, denoted as $\bsfv_{1,n}$. 

We combine the method in \cite{ParkEtAlDynamicSubarraysHybridPrecoding2017} with the Lorentzian mapping discussed in \cite{CarlsonEtAlHierarchicalCodebookDesignWith2023} to obtain a feasible DMA precoder. Let $\beeta_n = [\eta_{1,n}, \, \cdots, \, \eta_{\Ntuc, n}]$ be the waveguide attenuation coefficients for the $n$th DMA. The $n$th DMA beamformer can be written in terms of a vector $\bsfg_n$ with unit-modulus entries as 
\begin{equation}
  \label{eq:dma-beam-decomp}
  \bsff_{\dma,n} = \frac{\beeta_n \odot \bsfg_n - \beeta_n \odot \cj \bone}{2}.
\end{equation}
Lorentzian beamforming computes a feasbile DMA beamformer $\bsff_{\dma,n}^{\sf L}$ by mapping the phase of $\bsfv_{1,n}$ to $\bsfg_n$ as
\begin{equation}
  \label{eq:Lorentzian-mapping-opt}
  \bsff_{\dma,n}^{\sf L} = \frac{\beeta_n \odot e^{\cj \angle \bsfv_{1,n}} - \beeta_n \odot j\bone}{2}.
\end{equation}
The DMA precoder $\bsfF_{\dma}^\star = \blkdiag\left( \bsff_{\dma, 1}^{\sf L}, \, \cdots, \, \bsff_{\dma, \Nt}^{\sf L}\right)$. Given $\bsfF_{\dma}^\star$, the transmitter then applies the method in \cite{ParkEtAlDynamicSubarraysHybridPrecoding2017} to the channel $\bsfH[k] \bsfF_{\dma}^\star$ to obtain both $\Fana$ and $\Fdig[k]$ under the hybrid precoding constraints. While suboptimal, this approach serves as a useful performance baseline to gauge the benefits of tri-hybrid in terms of energy efficiency.

\subsection{Power consumption models}

The power consumption of a precoding architecture is determined by the RF components required for each signal processing block. We denote the power consumption of the transmit components as follows: $P_{\sf PA}$ for a single power amplifier, $P_{\sf DAC}$ for a digital-to-analog converter (DAC) $P_{\sf LO}$ for a local oscillator, $P_{\sf RF}$ for the combination of the mixer, low-pass filter and hybrid with buffer, $P_{\sf PS}$ for a phase shifter. We also denote $L_{\sf{PS}}$ and $L_{\sf{D}}(N_{a})$ as the component loss due to phase shifters and an $N_a$-way power divider. We incorporate power consumption and component loss model for each architecture based on the work in \cite{RibeiroEtAlEnergyEfficiencyMmwaveMassive2018}, summarized in Table~\ref{table: power consumption}. Note that we assume that the varactor diodes used in the DMAs consume negligible power. The main factors that vary the power consumption for each architecture are the number of RF chains and phase shifters. In the following, we compute the power consumption of various precoding architectures.

\begin{table}[!t]
    \centering
    \caption{Power consumption and component loss in the transmitter architecture model.}
    \begin{tabular}{|c|c|c|}
        \hline
        \textbf{Component} & \textbf{Notation} & \textbf{Value} \\ \hline
    
        Power amplifier efficiency & $\eta_{\sf{PA}}$ & 27\% \\
        Power amplifier & $P_{\sf{PA}}$ & $P_{\sf{PA}} = \frac{P_{\sf{T}}}{\eta_{\sf{PA}}}$ \\
        Phase shifter (active ; passive) & $P_{\sf{PS}}$ & 21.6 ; 0 mW \\
        DAC & $P_{\sf{DAC}}$ & 75.8 mW\\
        Local oscillator & $P_{\sf{LO}}$ & 22.5 mW \\
        RF chain & $P_{\sf{RF}}$ & 31.6 mW \\ 
        Varactor diode & $P_{\sf{VAR}}$ & 0 mW \\ \hline
        Phase shifter (active ; passive) & $L_{\sf{PS}}$ & $-$2.3 ; 8.8 dB \\
        Two-way power divider & $\bar{L}_{\sf{D}}$ & 0.6 dB \\ 
         \hline
        
    \end{tabular}
    \label{table: power consumption}
  \end{table}

\emph{Tri-hybrid}: The tri-hybrid architecture encompasses the system described in Section IV. Assuming a $\Ntrf$-way power divider and $\Nt$ phase shifters, the component loss for the tri-hybrid architecture is given as
\begin{equation}\label{eq: L TH}
    L_{\sf{TH}} = \Ld (\Ntrf) \Lps,
\end{equation}
and the power consumption as
\begin{equation}\label{eq: Pc TH}
    P_{\TH} = P_{\sf{LO}} +P_{\sf{PA}}+  N_{\sf{RF}} ( 2P_{\sf{DAC}}(b_{\sf{DAC}},F_s)+P_{\sf{RF}})+ \Nt P_{\sf{PS}}.
\end{equation}
The tri-hybrid case combines DMA precoding, analog phase shifters, and digital precoding to maximize performance while maintaining a low power consumption. The number of radiating elements is $\Ntrad = \Nt M \Ntuc$.

\emph{DMA-only}: The DMA-only (DO) architecture considers only DMA-based precoding with no phase shifters or digital precoding. DO precoding, therefore, contains only $1$ DMA with $1$ RF chain and $0$ phase shifters. 

Since there is only one RF chain, the power consumption of this architecture is given by
\begin{equation}\label{eq: Pc DA}
    P_{\sf{DO}} = P_{\sf{LO}} +P_{\sf{PA}}+  ( 2P_{\sf{DAC}}(b_{\sf{DAC}},F_s)+P_{\sf{RF}}).
\end{equation}
This architecture only has $\Ntrad = \Ntuc$ radiating elements, so its size is completely dictated by the DMA dimensions. Due to its extremely simple implementation, this architecture yields the lowest power consumption among all baselines.

\emph{DMA-digital hybrid}: The DMA-digital hybrid architecture (DH) combines DMA precoding with a digital precoder and does not use analog precoding. 
The DMA-hybrid architecture contains $\Ntrf$ RF chains and $0$ phase shifters, so the component loss for the $\Ntrf$-way power divider is
\begin{equation}\label{eq: L DH}
    L_{\DH} = \Ld (\Ntrf),
\end{equation}
and the power consumption is given by
\begin{equation}\label{eq: Pc DH}
    P_{\DH} = P_{\sf{LO}} +P_{\sf{PA}}+ \Ntrf ( 2P_{\sf{DAC}}(b_{\sf{DAC}},F_s)+P_{\sf{RF}}).
\end{equation}
We assume that the output of each RF chain only attaches to a single DMA, giving $\Ntrad = \Nt \Ntuc$. The addition of digital precoding provides better performance than the DMA-only architecture at the cost of increased power consumption.

\emph{Fully-analog}: The subsequent architectures use antenna arrays with typical non-reconfigurable elements with the same orientation as the tri-hybrid DMA arrays. In all following cases, we have $\Ntrad = \Nt$. We define the component loss for the $\Ntrf$-way power divider and $\Nt$ phase shifters as
\begin{equation}\label{eq: L ana}
    L_{\mathsf{A}} = \Ld (\Ntrf) \Lps,
\end{equation}
and the power consumption for a single RF chain as
\begin{equation}\label{eq: Pc ana}
    P_{\mathsf{A}} = P_{\sf{LO}} +P_{\sf{PA}}+  ( 2P_{\sf{DAC}}(b_{\sf{DAC}},F_s)+P_{\sf{RF}})+\Nt P_{\mathsf{PS}}.
\end{equation}
Compared to the DO architecture, the fully-analog system consumes more power but offers greater beamforming capabilities.

\emph{Partially-connected hybrid}: The hybrid partially-connected architecture (HP) considers both an analog and digital precoder. To reduce the number of RF chains, and subsequently the power consumption, the hybrid partially-connected architecture assumes the same subarray design as the DMA, where there are $\Ntrf$ RF chains. We assume each RF chain connects to a subarray consisting of $\Nsub = \Nt / \Ntrf$ antennas. The component loss for the $\Nsub$-way power divider is
\begin{equation}\label{eq: L HP}
    L_{\HP} = \Ld (\Nsub)\Lps,
\end{equation}
and the power consumption is given by
\begin{equation}\label{eq: Pc HP}
    P_{\HP} = P_{\sf{LO}} +P_{\sf{PA}}+ \Ntrf ( 2P_{\sf{DAC}}(b_{\sf{DAC}},F_s)+P_{\sf{RF}}) +\Nt P_{\sf{PS}}
\end{equation}
In terms of hybrid precoding, the partially-connected architecture achieves high energy efficiency as it significantly reduces the number of RF chains compared to a fully-connected architecture. A partially-connected hybrid precoding architecture also provides a key comparison with the tri-hybrid architecture, since both architectures contain subarrays of antenna elements.

\emph{Fully-connected hybrid}: The hybrid fully-connected architecture (HF) is similar to the partially-connected architecture, but instead of dividing the antenna array into subarrays, each antenna element is connected to all RF chains. Thus, there are a total of $\Nt$ phase shifters for the fully-connected architecture. The component loss for the $\Ntrf$-way power divider is
\begin{equation}\label{eq: L HF}
    L_{\HF} = \Ld (\Nt)\Lps\Lc(\Ntrf),
\end{equation}
and the power consumption is given by
\begin{equation}\label{eq: Pc HF}
    P_{\HF} = P_{\sf{LO}} +P_{\sf{PA}}+ \Ntrf ( 2P_{\sf{DAC}}(b_{\sf{DAC}},F_s)+P_{\sf{RF}}) + \Nt P_{\sf{PS}}
\end{equation}
Compared to the partially-connected architecture, the fully-connected architecture yields better performance since each antenna is connected to all RF chains, but results in higher power consumption from the added phase shifters.

\emph{Fully-digital architecture}: Lastly, we consider a fully-digital architecture. The fully-digital architecture assumes that each antenna element is connected to its own RF chain, which gives the greatest amount of precoding control but also significantly increases the power consumption compared to the hybrid precoding architectures. 
We consider no power divider in the digital architecture, leading to 
\begin{equation}\label{eq: L FD}
    L_{\FD} = 1,
\end{equation}
and the power consumption for the fully-digital case is given by
\begin{equation}\label{eq: Pc FD}
    P_{\FD} = P_{\sf{LO}} +P_{\sf{PA}}+ \Nt ( 2P_{\sf{DAC}}(b_{\sf{DAC}},F_s)+P_{\sf{RF}}).
\end{equation}
This architecture serves as the upper performance limit in terms of spectral efficiency.

\emph{Summary and comparisons}: We summarize the capabilities and power consumption of each of the different MIMO architectures in Table~\ref{table: notation}. A salient feature of the DMA-based architectures is the significant increase in number of radiating elements for the same power consumption as a hybrid architecture. Both the DMA-hybrid and the tri-hybrid architectures replace each antenna with at least one DMA, which also means the same number of radiating elements can be achieved with a lower power consumption. This benefit, of course, comes at the cost of the reduced flexibility of the DMA precoding matrix. 

The comparison above does not consider all possible types of architectures and their benefits. The types of analog and DMA components can be adjusted even within the considered implementations. For example, the choice of an active vs. a passive phase shifter can have an effect on the component loss and power consumption of analog precoding~\cite{RibeiroEtAlEnergyEfficiencyMmwaveMassive2018}. Similarly, DMA configuration can be achieved through PIN diodes instead of varactor diodes, which can simplify the hardware design \cite{ShlezingerEtAlDynamicMetasurfaceAntennas6g2021}. Other options for reducing the power consumption include reducing the resolution of data converters~\cite{AlkhateebEtAlMimoPrecodingAndCombining2014} and replacing RF precoding with lens arrays~\cite{LajosBeamspace2019}. The main purpose of our comparison is to establish that there are concrete advantages to the tri-hybrid architecture, but we leave the design of an optimal architecture for future work.

\begin{table*}[!t]
  \centering
  \caption{Notation used to represent various precoding architectures.}
  \begin{tabular}{|c|c|c|c|c|c|c|c|}
    \hline
    Abbr. & MIMO architecture & \begin{tabular}{@{}c@{}}RF \\ Chains\end{tabular} & \begin{tabular}{@{}c@{}}Phase \\ Shifters\end{tabular} & \begin{tabular}{@{}c@{}}Radiating \\ elements\end{tabular} & \begin{tabular}{@{}c@{}}Digital \\ precoding\end{tabular} & \begin{tabular}{@{}c@{}}Analog \\ precoding\end{tabular} & \begin{tabular}{@{}c@{}}DMA / EM \\ precoding\end{tabular} \\ \hline
    DO  & DMA-only & 1 & 0 & $\Ntuc$ & & & \checkmark \\ \hline
    DH  & DMA-digital hybrid & $\Ntrf$ & 0 & $\Nt \Ntuc$ & \checkmark & & \checkmark \\
    \hline
    TH & Tri-hybrid & $\Ntrf$ & $\Nt$ & $\Nt M \Ntuc$  & \checkmark & \checkmark & \checkmark \\
    \hline
    A  & Fully-analog & 1 & $\Nt$ &  $\Nt$ & \checkmark & \checkmark & \\ \hline
    HP  & Partially-connected hybrid & $\Ntrf$ & $\Nt$ & $\Nt$ &\checkmark & \checkmark & \\ \hline
    HF  & Fully-connected hybrid & $\Ntrf$ & $\Nt \Ntrf$ & $\Nt$ &\checkmark & \checkmark & \\ \hline
    FD  & Fully-digital & $\Nt$ & 0 & $\Nt$ &\checkmark &\checkmark &  \\ \hline
    
  \end{tabular}
  \label{table: notation}
\end{table*}

\subsection{Numerical results}

We compare various precoding architectures in terms of spectral efficiency and energy efficiency. We assume a multi-path channel with $L$ components. For the $\ell$th path, let $\trl$ be the elevation angle of arrival (AoA), $\prl$ be the azimuth AoA, $\ttl$ be the elevation angle of departure (AoD), and $\ptl$ be the azimuth AoD. Let $k_0 = \frac{2\pi}{\lambda}$ be the free-space wavenumber for a wavelength~$\lambda$. We assume the DMA as a UPA oriented in the xy plane, with $\Ntuc$ elements in the x direction and $M \Nt$ waveguides in the y direction. We denote the receive inter-element spacing as $\dr$ and the transmit inter-element spacing in the x- and y-axis as $\dx$ and $\dy$. The receive array steering vector is
\begin{align}
    \ar(\trl,\prl) &= \frac{1}{\sqrt{\Nr}} \Big[ e^{\sfj(0)k_0 \dr \sin\trl \cos \prl}, \ldots, \notag \\
    &\qquad e^{\sfj(\Nr-1)k_0 \dr \sin\trl \cos \prl} \Big]^T,
\end{align}
and the transmit UPA response vector is
\begin{align}
    \at&(\ttl,\ptl) = \nonumber\\
   &\frac{1}{\sqrt{\Ntrad}} \Bigr[ e^{\sfj(0)k_0 \dx \sin\trl \cos \prl}e^{\sfj(0)k_0 \dy \sin\trl \sin \prl} \nonumber \ldots \\ & \ldots, e^{\sfj(\Ntuc-1)k_0 \dx \sin\trl \cos \prl}e^{\sfj(M \Nt-1)k_0 \dy \sin\trl \sin \prl} \Bigr]^T.
\end{align}
For comparison purposes, we assume that transmit hybrid architectures without DMAs use an identical arrangement in which isotropic antennas are used in place of the unit cells.

We characterize the wireless channel through a wideband geometric model. Let $\al$ and $\taul$ represent the complex path gain and delay of the $\ell$th multi-path component. We assume a symbol duration $T_{\sf s}$. In the time-domain, the pulse-shaping filter for delay $d$ is denoted as $p(d T_{\sf s}- \taul)$. The frequency-domain filter response is then
\begin{equation}
    \omega_{\tau_{\ell}}[k] = \sum\limits_{d=0}^{D-1} p(dT_{\sf{s}}-\tau_{\ell}) e^{\frac{-j 2 \pi k d}{K}}.
\end{equation}
The wireless channel is modeled as the sum of the multi-path components as
\begin{equation}
    \bsfH[k] = \sqrt{\frac{\Ntrad \Nr}{L}}\sum\limits_{\ell=1}^L \al \wtau[k] \mathbf{a}_{\sf{r}}(\theta_{{\sf{r}},\ell},\phi_{{\sf{r}},\ell}) \mathbf{a}^*_{\sf{t}}(\theta_{{\sf{t}},\ell},\phi_{{\sf{t}},\ell}).
\end{equation}
We assume that the wireless channel is normalized such that $\mathbb{E}[||\bsfH[k]||_{\fro}^2] = \Ntrad \Nr$.

\begin{figure*}
    \centering
\begin{tabular}{cc}
\includegraphics[width = 0.45\textwidth]{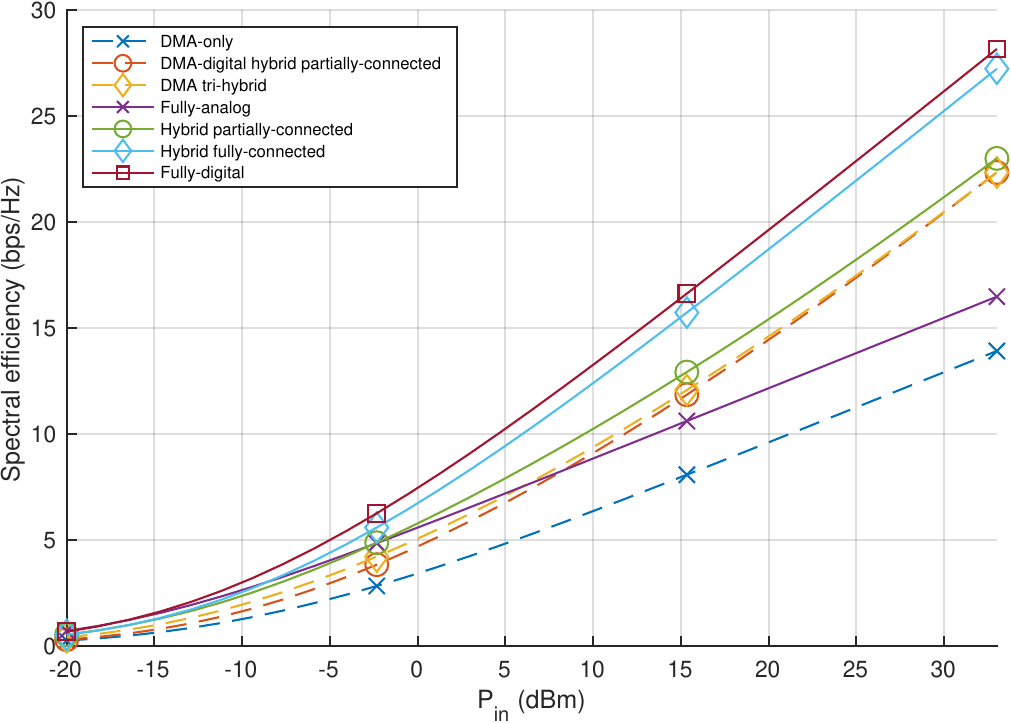} & \includegraphics[width = 0.45\textwidth]{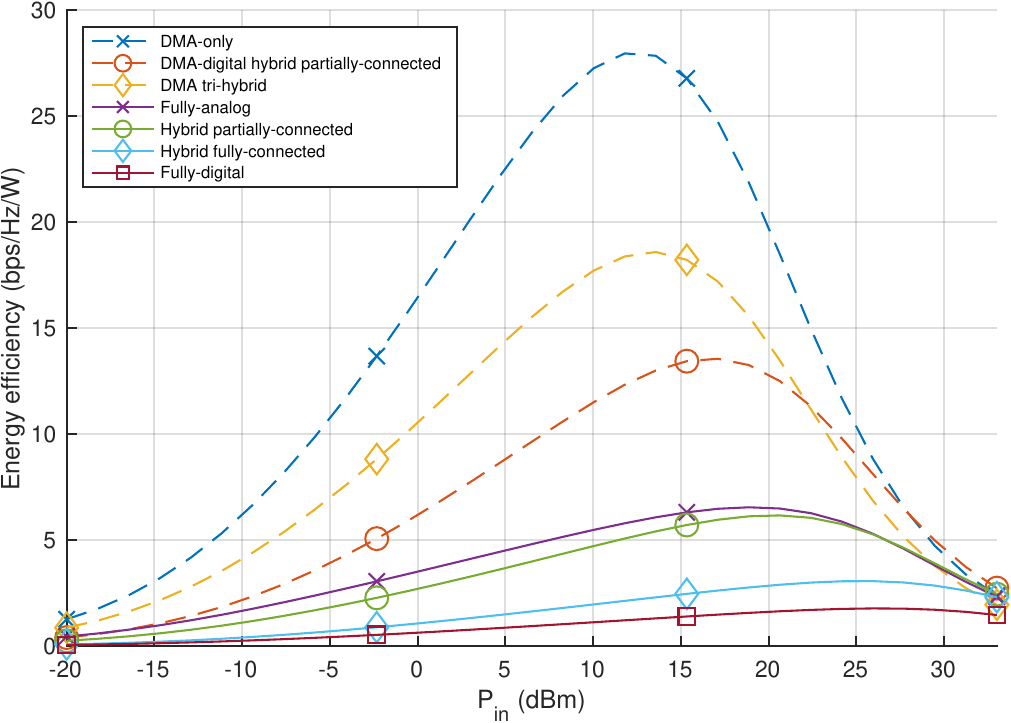} \\
(a) & (b)
\end{tabular}
\caption{Performance comparisons between the tri-hybrid architecture and baselines (fully-digital, fully-analog, and hybrid precoding) against input power. While the tri-hybrid precoding shows lower spectral efficiency than existing architectures in (a), the energy efficiency results in (b) demonstrate a good tradeoff between rate and power consumption.}
\label{fig_tri_hybrid_performance}
\end{figure*}

In Fig. \ref{fig_tri_hybrid_performance}, we show both the spectral efficiency and the energy efficiency of the DMA-based tri-hybrid architecture and other associated baselines. For an operating frequency of $15$ GHz, we set $K = 128$, $L=2$, $\Ns = 2$, $\Nr = 64$, $\dr = \lambda/2$. At the transmitter, we set $\dx = \lambda/4$, $\dy = \lambda/2$, and $\Ntrad = 64$. For the DMA-based hybrid architectures, we set $M = 2$, and $\Nt = \Ntrf = 2$. We also assume a raised-cosine filter with roll-off factor of $1$ and a symbol duration of $T_{\sf s} = 50$ ms. Given precoders $\Fdig[k]$, $\Fana$, and $\Fdma$, the spectral efficiency is computed as
\begin{equation}
    C = \sum_{k=1}^K \log_2 \abs{\bI_{\Nr} + \frac{1}{\sigma^2} \bsfH[k] \Fdma \Fhyb[k] \Fhybc[k] \Fdmac \bsfH^*[k]}.
\end{equation}
To ensure a fair comparison between different architectures, we assume each system has the same input power, rather than transmit power, and also the same number of radiating elements $\Ntrad$. Denoting the input power as $P_{\sf in}$ and architecture component loss as $L_{\sf c}$, the transmit power $P$ is
\begin{equation}
    P = \frac{P_{\sf in}}{L}.
\end{equation}
Further denoting the system power consumption as $P_{\sf cons}$, energy efficiency of the system is computed as the ratio of the spectral efficiency and the power consumption as $C/P_{\sf cons}$.

The results in Fig. \ref{fig_tri_hybrid_performance} demonstrate the key trade-offs of the tri-hybrid architecture. The spectral efficiency of the tri-hybrid architecture is lower than that of fully-digital and hybrid architectures as shown in Fig. \ref{fig_tri_hybrid_performance}(a). \edit{The main reason for the lower performance is the DMA precoding constraints and waveguide losses. The Lorentzian constraint couples the beamforming phase and amplitude in a way that results in lower magnitude weights compared to phase shifters. The DMA itself also suffers losses from waveguide attenuation as the transmit signal propagates through the antenna.} Despite these detriments, however, the tri-hybrid architecture outperforms architectures that rely solely on RF precoding. Fig.~\ref{fig_tri_hybrid_performance}(b) demonstrates that the true benefit of the tri-hybrid architecture lies in how it achieves a moderate spectral efficiency at very low power consumption. For a broad range of input power values, we see that all tri-hybrid architectures significantly outperform analog, hybrid, and digital architectures.

\section{Future outlook}
\label{sec_future}


The tri-hybrid MIMO architecture is the next evolution of the hybrid analog-digital MIMO architecture. In this paper, we introduced the tri-hybrid MIMO architecture as a combination of precoding in the antenna, analog and digital domains. We described several possible configurations of tri-hybrid MIMO at a higher level, and then went into more detail with a design that leverages DMAs for the reconfigurable antenna element. Using this specific design, we illustrated several points of distinction: a new layer of antenna precoding, different constraints on the antenna precoding matrices depending on the specific designed antennas, and novel in terms of spectral efficiency and energy consumption. With further research and development, this tri-hybrid architecture could play a pivotal role in the future of wireless communications, addressing the stringent requirements of 6G and beyond.

Beyond its performance gains, the tri-hybrid architecture offers a unifying framework that brings together previously proposed approaches in MIMO system design. Integrating antenna-based precoding with conventional analog and digital precoding provides a common ground for evaluating and comparing different approaches. These include dynamic scattering arrays with reconfigurable parasitic elements \cite{dardari2024DSA}, continuous apertures composed of subwavelength radiating elements \cite{DamicoHolographicMIMOCommunications2024}, liquid metal antennas \cite{BhaMaDic:RESHAPE:-A-Liquid-Metal-Based:21}, and arrays with movable elements \cite{DoEtAlReconfigurableUlasLineSight2021}. This cohesive perspective streamlines the design of new tri-hybrid architectures that combine the benefits of different approaches in the literature. While the proposed architecture demonstrates energy efficiency and enables signal processing through the antennas, there are many key areas that require further research and innovation. Additionally, waveguide-fed metasurfaces and dynamically reconfigurable DMA architectures can be seen as specialized forms of tri-hybrid MIMO, where electromagnetic precoding is implemented through tunable metasurface elements \cite{SmithEtAlAnalysisWaveguideFedMetasurface2017, CarlsonEtAlHierarchicalCodebookDesignWith2023}. Similarly, hybrid beamforming architectures incorporating  reconfigurable reflectarrays \cite{hum2014reconfigurable} or reconfigurable transmitarray \cite{SIMcomputing_1720551601} provide alternative pathways to achieving reconfigurable capabilities within the tri-hybrid framework.

\emph{Advanced antenna designs:} Reconfigurable antennas form the backbone of the tri-hybrid MIMO architecture. But many current designs are still limited in terms of their adaptability and efficiency. Future work should explore the integration of novel materials, such as metasurfaces and tunable metamaterials, to enable finer control over antenna parameters like beamforming, polarization, and frequency response. Work is also needed to evaluate these designs as part of a larger tri-hybrid precoding design, and not just in terms of traditional antenna metrics like impedance parameters, gain patterns, directivity and efficiency \cite{StutzmanThieleAntennaTheoryAndDesign2012}. These innovations could unlock higher degrees of reconfigurability and improved performance in dynamic environments.

\emph{Multi-parameter reconfigurability:} A significant challenge lies in developing antennas that can reconfigure multiple parameters simultaneously (e.g., polarization, frequency, and gain pattern). Multi-parameter reconfigurability could greatly enhance system flexibility but introduces increased design complexity. Balancing this complexity with practical hardware and software solutions will be crucial for the future success of tri-hybrid MIMO systems.

\emph{Circuit and hardware considerations:}
The implementation of low-power, high-efficiency RF chains, phase shifters, and other analog components remains a challenge, particularly as systems scale. Future research should focus on optimizing hardware designs to further reduce power consumption while maintaining performance. Additionally, scalable manufacturing processes that ensure cost-effectiveness will be critical for large-scale deployment.

\emph{Precoder design:}
New algorithms are needed to optimize the antenna precoder. Reconfigurable antennas introduce a third phase of precoding compared to hybrid analog-digital methods with its own set of constraints. New approaches are needed to formulate meaningful objective functions and to solve them to acquire each of the three layers of precoding. For example, one promising direction might be to solve for the ideal unconstrained precoder then solve a nearness problem to find the best tri-hybrid precoder in close proximity. This approach was used in early papers on hybrid precoding \cite{O.E.AyachEtAlSpatiallySparsePrecodingMillimeter2014}, but exploiting certain array geometries that are not necessarily available when reconfigurable antennas are used. Nonlinear optimization tools and optimization via machine learning may useful in devising new precoding methods. 

\emph{Channel estimation and tracking:}
The beam-based MIMO design in 5G was inspired to a large extent by the challenges associated with configuring hybrid analog/digital MIMO links \cite{DreifuerstHeathMassiveMIMO5GHow2023}. Beam-based MIMO can be used in a first phase to solve for the analog precoding via sweeping a broad set of beams, with a possible refinement stage, followed by estimating the effective channel and digital precoding. It is unclear how to incorporate the antenna configuration? Should it also be done as part of an initial sweeping procedure? Or should new methods for compressive channel estimation be developed like those in \cite{J.Rodriguez-FernandezEtAlFrequencyDomainCompressiveChannelEstimation2018} but where sensing antenna/analog/digital precoders are used to make measurements and estimate the high dimensional MIMO channel. There are other challenges as well, such as how to design the training sequences and how to track and update the tri-hybrid precoders once they are found.


\bibliographystyle{IEEEtran}
\bibliography{refs_rc}

\balance

\begin{IEEEbiography}
[{\includegraphics[width=1in,height=1.25in,clip,keepaspectratio]{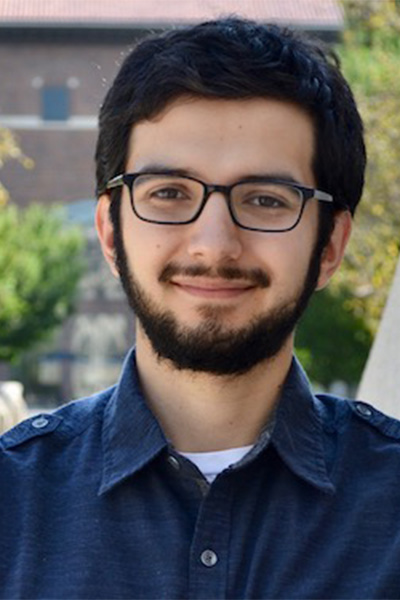}}]{Miguel Rodrigo Castellanos} (Member, IEEE) 
received the B.S. (Hons.) and Ph.D. degrees in electrical engineering from Purdue University, West Lafayette, IN, USA, in 2015 and 2020, respectively. From 2020 to 2021, he was with The University of Texas at Austin. Since 2021, he has been with North Carolina State University and is currently an Assistant Professor in the Department of Electrical Engineering and Computer Science at the University of Tennessee, Knoxville, TN, USA. During the summers of 2017 and 2019, he was a Research Intern at Qualcomm Flarion Technologies, Bridgewater, NJ, USA, and Nokia Networks, Naperville, IL, USA. He is currently a Post-Doctoral Researcher.  

His research interests include massive MIMO, near-field communication, antenna theory, exposure-constrained communication, and signal processing. He coauthored an article that won the 2022 Fred W. Ellersick MILCOM Best Paper Award and the 2022 Vanu Bose Best Paper Award.
\end{IEEEbiography}

\begin{IEEEbiography}[{\includegraphics[width=1in,height=1.25in,clip,keepaspectratio]{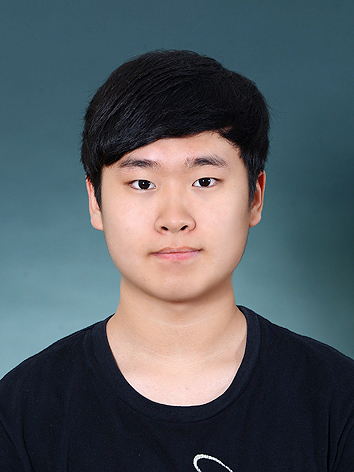}}]{Siyun Yang} (Graduate Student Member, IEEE) received the B.S. degree from the School of Integrated Technology, Yonsei University, South Korea, in 2022, where he is currently pursuing his Ph.D. degree. His research interests include multiple-input multiple-output (MIMO) technology for B5G/sixth-generation (6G) communications such as  deep learning based precoder design, integrated sensing and communications (ISACs), and full-duplex (FD) radios. He is a co-recipient of a Bronze Medal and two Honor Medals from the Humantech Paper Contest in 2024 and 2025.

\end{IEEEbiography}

\begin{IEEEbiography}[{\includegraphics[width=1in,height=1.25in,clip,keepaspectratio]{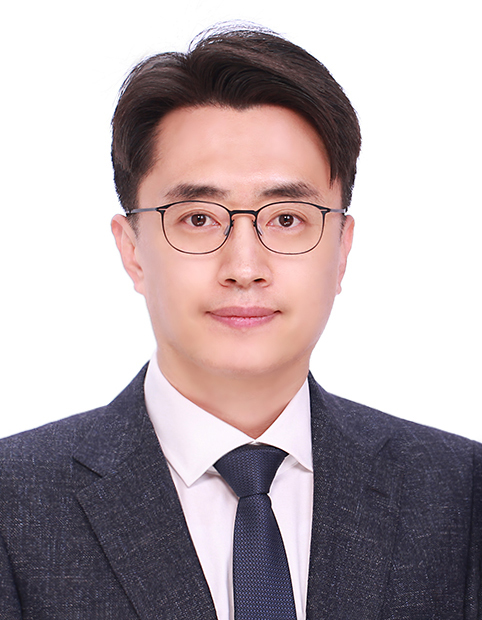}}]{Chan-Byoung Chae} (Fellow, IEEE) received the Ph.D. degree in electrical and computer engineering from The University of Texas at Austin (UT), USA in 2008, where he was a member of wireless networking and communications group (WNCG). Prior to joining UT, he was a Research Engineer at the Telecommunications R\&D Center, Samsung Electronics, Suwon, South Korea, from 2001 to 2005. He is currently an Underwood Distinguished Professor and Lee Youn Jae Fellow (Endowed Chair Professor) with the School of Integrated Technology, Yonsei University, South Korea. Before joining Yonsei, he was with Bell Labs, Alcatel-Lucent, Murray Hill, NJ, USA, from 2009 to 2011, as a Member of Technical Staff, and Harvard University, Cambridge, MA, USA, from 2008 to 2009, as a Post-Doc. Fellow and Lecturer. 

Dr. Chae was a recipient/co-recipient of the Ministry of Education Award in 2024, the KICS Haedong Scholar Award in 2023, the CES Innovation Award in 2023, the IEEE ICC Best Demo Award in 2022, the IEEE WCNC Best Demo Award in 2020, the Best Young Engineer Award from the National Academy of Engineering of Korea (NAEK) in 2019, the IEEE DySPAN Best Demo Award in 2018, the IEEE/KICS Journal of Communications and Networks Best Paper Award in 2018, the IEEE INFOCOM Best Demo Award in 2015, the IEIE/IEEE Joint Award for Young IT Engineer of the Year in 2014, the KICS Haedong Young Scholar Award in 2013, the IEEE Signal Processing Magazine Best Paper Award in 2013, the IEEE ComSoc AP Outstanding Young Researcher Award in 2012, and the IEEE VTS Dan. E. Noble Fellowship Award in 2008. 

Dr. Chae has held several editorial positions, including Editor-in-Chief of the IEEE Transactions on Molecular, Biological, and Multi-Scale Communications, Senior Editor of the IEEE Wireless Communications Letters, and Editor of the IEEE Communications Magazine, IEEE Transactions on Wireless Communications, and IEEE Wireless Communications Letters. He was an IEEE ComSoc Distinguished Lecturer from 2020 to 2023 and is an IEEE VTS Distinguished Lecturer from 2024 to 2025. He is an elected member of the National Academy of Engineering of Korea.
\end{IEEEbiography}

\begin{IEEEbiography}[{\includegraphics[width=1in,height=1.25in,clip,keepaspectratio]{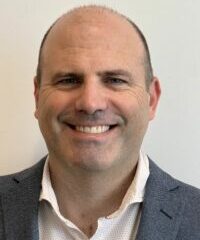}}]{Robert W. Heath, Jr.} (Fellow, IEEE) is currently the Charles Lee Powell Chair of Wireless Communications with the Department of Electrical and Computer Engineering, University of California at San Diego, CA, USA. He received the B.S. and M.S. degrees from the University of Virginia, Charlottesville, VA, USA in 1996 and 1997 respectively, and the Ph.D. from Stanford University, Stanford, CA, USA in 2002, all in electrical engineering. He is also the President and the CEO of MIMO Wireless Inc.

From 2002 to 2020, he was with The University of Texas at Austin, most recently as the Cockrell Family Regents Chair of Engineering and the
Director of UT SAVES. From 2020 to 2023, he was the Lampe Distinguished Professor with North Carolina State University and the Co-Founder of 6GNC. He has authored \emph{Introduction to Wireless Digital Communication} (Prentice Hall, 2017) and \emph{Digital Wireless Communication: Physical Layer Exploration Lab Using the NI USRP} (National Technology and Science Press, 2012) and coauthored \emph{Millimeter Wave Wireless Communications} (Prentice Hall, 2014) and \emph{Foundations of MIMO Communication} (Cambridge University Press, 2018). 

He has coauthored numerous award-winning conference and journal articles, including the 2017 Marconi Prize Paper Award, the 2019 IEEE Communications Society Stephen O. Rice Prize, the 2020 IEEE Signal Processing Society Donald G. Fink Overview Paper Award, the 2021 IEEE Vehicular Technology Society Neal Shepherd Memorial Best Propagation Paper Award, and the 2022 IEEE Vehicular Technology Society Best Vehicular Electronics Paper Award. His other notable honors include the 2017 EURASIP Technical Achievement Award, the 2019 IEEE Kiyo Tomiyasu Award, the 2021 IEEE Vehicular Technology Society James Evans Avant Garde Award, and 2025 IEEE/RSE James Maxwell Medal. In 2017, he was elected a Fellow of the National Academy of Inventors. He served as a Member-at-Large on the IEEE Communications Society Board of Governors (2020–2022) and the IEEE Signal Processing Society Board of Governors (2016–2018). From 2018 to 2020, he was the Editor-in-Chief of IEEE Signal Processing Magazine.

Beyond his academic and professional contributions, he is a licensed Amateur Radio Operator, a Private Pilot, and a registered Professional Engineer in Texas. He is also an elected member of the National Academy of Engineering. 
\end{IEEEbiography}

\end{document}